\documentclass[11pt,epsf]{article}
\usepackage{amsmath}
\usepackage{amsfonts}
\usepackage{amssymb}
\usepackage{graphicx}
\newcommand {\dfn} {\stackrel{\Delta} {=}}
\newcommand {\exe} {\stackrel{\cdot} {=}}
\newcommand {\lexe} {\stackrel{\cdot} {\le}}
\newcommand {\gexe} {\stackrel{\cdot} {\ge}}

\newcommand {\bx} {\mbox{\boldmath $x$}}
\newcommand {\hbx} {\hat{\mbox{\boldmath $x$}}}
\newcommand {\tbx} {\tilde{\mbox{\boldmath $x$}}}

\newcommand {\by} {\mbox{\boldmath $y$}}

\newcommand {\bE} {\mbox{\boldmath $E$}}

\newcommand {\bX} {\mbox{\boldmath $X$}}

\newcommand{\calA}{{\cal A}}

\newcommand{\calC}{{\cal C}}

\newcommand{\calE}{{\cal E}}

\newcommand{\calI}{{\cal I}}

\newcommand{\calX}{{\cal X}}
\newcommand{\calY}{{\cal Y}}

\begin{document}
\thispagestyle{empty}
\title{The Generalized Random Energy Model and\\
its Application to the Statistical Physics of Ensembles
of Hierarchical Codes
}
\author{Neri Merhav
}
\date{}
\maketitle

\begin{center}
Department of Electrical Engineering \\
Technion - Israel Institute of Technology \\
Haifa 32000, ISRAEL \\
\end{center}
\vspace{1.5\baselineskip}
\setlength{\baselineskip}{1.5\baselineskip}

\begin{abstract}
In an earlier work, the statistical physics 
associated with finite--temperature decoding
of code ensembles, along with the 
relation to their random coding error exponents, were explored 
in a framework that is analogous to Derrida's random energy model (REM)
of spin glasses, according to which the energy levels of the various
spin configurations are independent random variables.
The generalized REM (GREM) extends the REM in that it
introduces correlations between energy levels in an hierarchical structure.
In this paper, we explore some analogies 
between the behavior of the GREM and that of
code ensembles which have parallel hierarchical structures. In particular,
in analogy to the fact that the GREM may have different types of phase
transition effects, depending on the parameters of the model, then the
above--mentioned hierarchical code ensembles behave substantially differently
in the various domains of the design parameters of these codes. We make an
attempt to explore the insights that can be imported from the statistical mechanics of the GREM 
and be harnessed to serve for code design considerations and guidelines.\\
\\

\end{abstract}

\section{Introduction}

In the last few decades it has become apparent that many problems in
Information Theory have analogies to certain problems in the
area of statistical physics of disordered systems.
Such analogies are useful because physical insights, as well as
statistical mechanical tools and analysis techniques 
can be harnessed in order to advance the knowledge and the
understanding with regard to
the information--theoretic problem under discussion.

One important example of such an analogy is between the statistical physics
of disordered magnetic materials, a.k.a.\ spin glasses, and the behavior of
certain ensembles of random codes for source coding 
(see, e.g., \cite{HK05},
\cite{Murayama02},
\cite{KH05},
\cite{Tadaki07})
and for channel coding
(see, e.g., \cite{MM06} and references therein,
\cite{KanS99},
\cite{PS99},
\cite{Sourlas89},
\cite{Sourlas94},
\cite{KSNS01},
\cite{KabS99}
\cite{SK06},
\cite{MR06},
\cite{Montanari05a},
\cite{FLMR02},
\cite{MU07},
\cite{Montanari01},
\cite{Rujan93},
\cite{Montanari05b},
\cite{DW99}). 

Among the various models of interaction disorder 
in spin glasses, one of the most fascinating models is the {\it random
energy model} (REM), invented by Derrida in 
the early eighties \cite{Derrida80}, 
\cite{Derrida80b}, \cite{Derrida81} (see also, e.g.,
\cite{DW99},
\cite{CFP98},
\cite{Jana06},
for later developments). The REM
is on the one hand, extremely
simple and easy to analyze, and on the other hand, rich enough to exhibit
phase transitions. According to the REM, the different spin configurations
are distributed according to the Boltzmann distribution, 
namely, their probabilities are proportional to an exponential function of
their negative energies, but the configuration energies themselves are i.i.d.\ random
variables, hence the name random energy model.\footnote{More details on this and other
terminology described in the remaining part of this Introduction, will be given in
the Section \ref{bg}.}

In \cite[Chap.\ 6]{MM06}, M\'ezard and Montanari draw an interesting analogy between
the REM and the statistical physics pertaining to 
{\it finite temperature decoding} \cite{Rujan93} of ensembles of random block codes.
The relevance of the REM here is due to the fact that in this context, the
partition function that naturally arises has the log--likelihood function 
(of the channel output given the input codeword) as its energy function (Hamiltonian),
and since the codewords are selected at random, 
then the induced energy levels are random variables.
Consequently, the phase transitions of the REM are `inherited' by ensembles of random block codes,
as is shown in \cite{MM06}.
In \cite{Merhav07}, this subject was further studied and the free energies corresponding to
the various phases 
were related to random coding exponents of the probability of error
at rates below capacity and to the probability of correct decoding at rates above capacity.

While the REM is a very simple and interesting model for capturing disorder, as described above,
it is not quite faithful for the description of 
a real physical system. The reason is that according to
the REM, any two distinct spin configurations, no matter how similar and close to each other, have
independent, and hence unrelated, energies.
A more realistic model
must take into account the geometry and the structure of the physical system 
and thus allow dependencies between energies associated with closely 
related configurations. 

This observation has motivated Derrida to develop the {\it generalized
random energy model} (GREM) 
\cite{Derrida85}
(see also, e.g.,
\cite{DG86a},
\cite{DG86b},
\cite{DD01},
\cite{Saaskian97},
\cite{BK04},
\cite{BK02},
for later related work). The GREM extends the REM in that it introduces an
hierarchical structure in the form of a tree, 
by grouping subsets of (neighboring) spin configurations
in several levels, where the leaves of this tree correspond to the various configurations.
According to the GREM, for every branch in this tree, there is an associated 
independent randomly chosen
energy component. The total energy of each 
configuration is then the sum of these energy components
along the branches that form the path from the root of the tree to the leaf corresponding
to this configuration. This way, the degree of dependency between the energies of two
different configurations depends on the `distance' between them on the tree: More
precisely, it depends on
the number of common branches shared by their paths from the root up to the node at which their
paths split. The GREM is somewhat more complicated to analyze than the REM, but not substantially so.
It turns out that the number of phase transitions in the GREM depends on the parameters of the
model. If the tree has $k$ levels, there can be up to $k$ phase transitions, but there can also
be a smaller number. For example, in the case $k=2$, under a certain condition, there is only one
phase transition and the behavior of the free energy
in both phases is
just like in the ordinary REM.

In analogy to the above described relationship between the REM and the statistical
physics of random block codes, the natural question 
that now arises is whether the GREM and its phase transitions can give us some
insights about the behavior of code ensembles with some hierarchical structure
(e.g., tree--structured codes, successive refinement codes, etc.).
In particular, in what way do these phase transitions guide
us in the choice of the design parameters of these codes?
It is the purpose of this paper to explore these questions and to give at least some partial
answers.

We demonstrate that there is indeed an intimate relationship between the GREM
and certain ensembles of hierarchical codes. Consider, for example, a two--stage rate--distortion code
of block length $n=n_1+n_2$, where the first $n_1$ 
components of the reproduction vector, at rate $R_1$, depend
only on the first $n_1R_1$ bits of the compressed bitstream, 
and the last $n_2$ symbols of the reproduction
codeword, at rate $R_2$, depend on the entire bitstream of length $n_1R_1+n_2R_2$.
The overall rate of this code is, of course, the weighted average of $R_1$ and $R_2$
with weights proportional to $n_1$ and $n_2$, respectively. An ensemble of codes with this
structure is defined as follows: First, we randomly draw a rate $R_1$ codebook
of block length $n_1$ according to some distribution. 
Then, for each resulting codeword of length $n_1$, we randomly draw a rate $R_2$ codebook
of block length $n_2$.\footnote{Note 
that this is different from using the same second--stage
codebook for all first--part codewords,
in which case, this is just a combination 
two codebooks of length $n_1$ and $n_2$, operating
independently.}
Thus, the code has a tree structure with two 
levels, like a two--level GREM. The overall distortion 
of the code along the entire
$n$ symbols is the sum of partial 
distortions along the two segments, in analogy to
the above described additivity of the partial energies along 
the branches of the tree pertaining to the GREM, and since the codewords are
random, then so are the distortions they induce.

The motivation for this
class of codes, especially when the idea is generalized from two parts to a larger number
of $k$ parts, say, of equal length ($n_1=n_2=\ldots =n_k=n/k$), is that the delay, at least at the
decoder, is reduced from $n$ to $n/k$, because the decoder is causal in the level of segments
of length $n/k$. The following questions now arise:
Is there any inherent penalty, in terms of performance, 
for this ensemble of reduced delay
decoding codes? If so, how can we minimize this penalty? If not, how should we choose the design
parameters (i.e., $n_i$ and $R_i$, $i=1,\ldots,k$, for a given overall average rate $R$)
such that this code will `behave' like a full block code of length $n$?

For simplicity, let us return to the case $k=2$. For a given $R$ and $n$, we have two degrees
of freedom: the choices of $R_1$ and $n_1$ (which will then dictate $R_2$ and
$n_2$). Is it better to choose $R_1 > R_2$ or $R_1\le R_2$, if at all it makes any difference?
A similar question can be asked concerning $n_1$ and $n_2$. The answer depends, of course,
on our figure of merit. Obviously, if one is interested only in the asymptotic distortion,
the question becomes uninteresting, because then by choosing two independent codes\footnote{c.f.\
footnote no.\ 2.} for the
two parts, both at rate $R$, the overall distortion will be given by the distortion--rate
function, $D(R)$, just like that of 
the full unstructured code. For a given $n$, of course,
the redundancies will correspond to the shorter blocks $n_1$ and $n_2$, but this is a second
order effect. Here, we choose to examine performance in terms of the characteristic function of
the overall distortion, $\bE[\exp\{-s\cdot\mbox{distortion}\}]$.
This is, of course, a much more informative figure of merit than
the average distortion, because in principle, it gives information on the entire probability
distribution of the distortion. In particular, it generates all the moments of the distortion
by taking derivatives, and it is useful in deriving Chernoff bounds on probabilities of
large deviations events concerning the distortion. In the context of the analogy 
with statistical physics
and the GREM, this characteristic function can easily be related to the partition function
whose Hamiltonian is given by the distortion.

It turns out that the characteristic 
function of the distortion behaves in a rather
surprisingly interesting manner and 
with a direct relation to the GREM. For $R_1 < R_2$,
when the corresponding GREM has $k=2$ phase transitions, the characteristic function of
the distortion behaves like that of two independent block codes of lengths $n_1$ and $n_2$
and rates $R_1$ and $R_2$, thus the dependency between the two parts of the code is not exploited in terms
of performance. For $R_1 > R_2$, which is the case where the analogous GREM has only one phase
transition (and behaves exactly like the ordinary REM, which
is parallel to an ordinary random block code
with no structure),
the characteristic function behaves like that of a full
unstructured optimum block code at rate $R$ across a certain interval 
of small $s$,
but beyond a certain point,
it becomes inferior to that of a full code. For $R_1=R_2=R$, it behaves like
the unstructured code for the {\it entire} range of $s\ge 0$, 
but then one might as well use two independent block codes (and reduce the search complexity
at the encoder from $e^{nR}$ to $2e^{nR/2}$). The choices of $n_1$ 
and $n_2$ are immaterial in that sense, as long as they both grow linearly with $n$.
Thus, the conclusion is that it is best to use $R_1=R_2$, but if communication protocol
constraints dictate different rates at different segments,\footnote{
For example, this can be the case if there are additional users in the system
and the bandwidth allocation for each user changes in a dynamical manner, or if different
parts of the encoded information are transmitted via separate links with different capacities.}
then performance is better when $R_1 > R_2$ than when $R_1 < R_2$.
These results can be extended to the
case of $k$ stages.

A parallel analysis can be applied to analogous ensembles 
of (reduced delay) channel encoders 
of block length $n=n_1+n_2$ (for the case $k=2$), which have a similar tree
structure: Here, the first $n_1$ channel letters of each block depend
only on the first $n_1R_1$ information bits, whereas the other 
$n_2$ channel symbols depend on the entire information vector of
length $n_1R_1+n_2R_2$. The random codebook is again drawn
hierarchically in the same manner 
as before. If the code performance is judged in terms
of the error exponent, then once again, 
the choice $R_1\ge R_2$ is always better than
the choice $R_1< R_2$. Here, unlike the source coding problem,
there is an additional consideration: There are two types of incorrect
codewords that are competing with the correct one in the decoding process:
those for which the first $n_1$ channel inputs agree with those of the
correct codeword (the first segment is the same)
and those for which this is not the case. In this case, $R_2$ has to be
chosen sufficiently small so that the error term contributed by erroneous
codewords of the first kind would not dominate the probability of error. 
Considering the case $n_1=n_2=n/2$,
if the overall average rate is not too small, it is possible to choose
$R_1$ and $R_2$ so that the error exponent of this ensemble of codes
is not worse than that of an ordinary 
random code with no structure. This idea can be
extended to $k$ stages in a straightforward manner. In fact,
we propose a systematic procedure to allocate rates to the different stages in a way
that guarantees that the error exponent would be at least as good as
that of the classical random coding error exponent pertaining to an ordinary
random code at rate $R$.

The outline of this paper is as follows. 
In Section 2, a few notation conventions are described.
In Section 3, we provide
some more detailed background in statistical physics, with emphasis on the REM
and the GREM. 
Finally, in Section 4, we present our main results 
on hierarchical code ensembles of the type described
above, along with their relationship to the GREM.
Readers who are not interested in the relationship
with statistical physics (although this is one of the main points in the
paper) may skip Section 3 
and ignore, in Section 4, the comments on 
the statistical mechanical aspects, all this without
essential loss of continuity. 

\section{Notation Conventions}

Throughout this paper, scalar random
variables (RV's) will be denoted by capital
letters, like $X$ and $Y$, their sample values will be denoted by
the respective lower case letters, and their alphabets will be denoted
by the respective calligraphic letters.
A similar convention will apply to
random vectors and their sample values,
which will be denoted with the same symbols in the boldface font.
Thus, for example, $\bX$ will denote a random $n$-vector $(X_1,\ldots,X_n)$,
and $\bx=(x_1,...,x_n)$ is a specific vector value in $\calX^n$,
the $n$-th Cartesian power of $\calX$.

Sources and channels will be denoted generically by the letters $P$ and $Q$.
Specific letter probabilities corresponding to a source $Q$ will be
denoted by the corresponding lower case letters, e.g., $q(x)$ is the
probability of a letter $x\in\calX$. A similar convention will be applied
to the channel $P$ and the
corresponding transition probabilities, $p(y|x)$,
$x\in\calX$, $y\in\calY$. The expectation operator will be
denoted by $\bE\{\cdot\}$.

The cardinality of a finite set $\calA$ will be denoted by $|\calA|$.
For two positive sequences $\{a_n\}$ and $\{b_n\}$, the notation
$a_n\exe b_n$ means that $a_n$ and $b_n$ are asymptotically of the same
exponential order, that is, $\lim_{n\to\infty}\frac{1}{n}\ln\frac{a_n}{b_n}
=0$. Similarly, $a_n\lexe b_n$ means that $\limsup_{n\to\infty}\frac{1}{n}\ln\frac{a_n}{b_n}\le 0$,
etc. Information theoretic quantities like entropies and mutual
informations will be denoted following the usual conventions
of the Information Theory literature.

\section{Background}
\label{bg}

In this section, we provide some basic background in statistical physics,
focusing primarily on the REM, along with its relevance to ordinary ensembles of 
source and channel block codes, and then we extend the scope to the GREM. 

\subsection{General}
\label{general}

Consider a physical system with a large number $n$ of particles, 
which can be in a variety of `microstates' pertaining to the various combinations
of the microscopic physical states (characterized by position, momentum, spin, etc.) that 
these particles may have. For each such microstate of the system, which we shall
designate by a vector $\bx$, there is an 
associated energy, given by an energy function (Hamiltonian)
$\calE(\bx)$.
One of the most
fundamental results in statistical physics (based on the law of energy conservation and
the basic postulate that all microstates of the same energy level are equiprobable)
is that when the system is in equilibrium, the probability of a microstate $\bx$ is
given by the Boltzmann distribution
\begin{equation}
\label{bd}
P(\bx)=\frac{e^{-\beta\calE(\bx)}}{Z(\beta)}
\end{equation}
where $\beta$ is the inverse temperature, that is, $\beta=1/T$, $T$ being temperature,\footnote{
More precisely, $\beta=1/(kT)$, where $k$ is Boltzmann's constant, but following the common
abuse of the notation, we redefine $T\leftarrow kT$ as temperature (in units of energy).} and
$Z(\beta)$ is the normalization constant, called the {\it partition function}, which
is given by
$$Z(\beta)=\sum_{\bx} e^{-\beta\calE(\bx)}$$
or
$$Z(\beta)=\int d\bx e^{-\beta\calE(\bx)},$$
depending on whether $\bx$ is discrete or continuous. The role
of the partition function is by far deeper than just being a normalization factor, as
it is actually the key quantity from which many macroscopic physical quantities can be derived,
for example, the free energy is $F=-\frac{1}{\beta}\ln Z(\beta)$, the average internal
energy (i.e., the expectation of $\calE(\bx)$ where $\bx$ drawn is according (\ref{bd}))
is given by the negative derivative of $\ln Z(\beta)$, the heat capacity is obtained from 
the second derivative, etc.

One of the important examples of such a multi--particle physical system is that of a magnetic
material, in which each molecule has a magnetic moment, a three--dimensional vector which
tends to align with the magnetic field felt by that molecule. In addition to the 
influence of a possible external magnetic
field, there is also an effect of mutual interactions between the magnetic moments of
various (neighboring) molecules. Quantum mechanical considerations dictate that the set of possible
configurations of each magnetic moment (spin) is discrete: in the simplest case, it has only
two possible values, which we shall designate by $+1$ (spin up) and $-1$ (spin down). Thus, a spin
configuration, i.e., the vector of spins of $n$ molecules,
is designated by a binary vector $\bx=(x_1,\ldots,x_n)$, where each
component $x_i$ takes values in $\{-1,+1\}$ according to the spin of the $i$--th molecule,
$i=1,2,\ldots,n$. When the spins of a certain magnetic material tend to align in the
same direction, the material is called {\it ferromagnetic}, and a customary model of the 
Hamiltonian, the {\it Ising model}, is given by
\begin{equation}
\label{ham}
\calE(\bx)=-J\sum_{i,j}x_ix_j-B\sum_{i=1}^nx_i
\end{equation}
where the in first term, pertaining to the interaction,
$J > 0$ describes the intensity of the interaction with the summation
being defined over pairs of neighboring spins (depending on the geometry of the problem),
and the second term is associated with an external magnetic field (proportional to) $B$. 
When $J < 0$, the material is {\it antiferromagnetic}, namely, neighboring spins `prefer'
to be antiparallel. More general models allow interactions not only with immediate neighbors,
but also more distant ones, and then there are different strengths of interaction, depending
on the distance between the two spins. In this case, the first term is replaced, by the more
general form $-\sum_{i,j}J_{ij}x_ix_j$, where now the sum can be defined over all possible
pairs $\{(i,j)\}$.\footnote{Moreover, the interaction term may be generalized 
to include also summations over
triples of spins, quadruples, etc., but we will limit the discussion to pairs.}
Here, in addition to the ferromagnetic case, where all $J_{ij} > 0$, and the
antiferromagnetic case, where all $J_{ij} < 0$, there is also a situation where
some $J_{ij}$ are positive and others are negative, which is the case if a {\it spin glass}.
Here, not all spin pairs can be in their preferred mutual position (parallel/antiparallel),
thus the system may be {\it frustrated.}

To model situations of disorder, it is common to model $J_{ij}$ as random variables (RV's) with,
say, equal probabilities of being positive or negative. For example, 
in the Edwards--Anderson (EA) model \cite{EA75},
$J_{ij}$  are taken to be i.i.d.\ zero--mean Gaussian RV's when $i$ and $j$ are neighbors
and zero otherwise. In the Sherrington--Kirkpatrick (SK) model \cite{SK75}, all $\{J_{ij}\}$ are
i.i.d.\ zero--mean Gaussian RV's. Thus, the system has two levels of randomness: the
randomness of the interaction coefficients and the randomness of the spin configuration given
the interaction coefficients, according to the Boltzmann distribution. However, the two
sets of RV's are normally treated differently. The random coefficients are considered {\it
quenched} RV's in the terminology of physicists, 
namely, they are considered fixed in the time scale at which the spin 
configuration may vary. This is analogous to the situation 
of coded communication in a random coding paradigm: A randomly
drawn code should normally be thought of as a quenched 
entity, as opposed to the randomness of the source and/or the channel.

\subsection{The REM}
\label{rem}

In \cite{Derrida80},\cite{Derrida80b},\cite{Derrida81}, Derrida took the above described 
idea of randomizing the (parameters of the) Hamiltonian to an extreme, and suggested a model of spin
glass with disorder under which the energy levels $\{\calE(\bx)\}$ are simply i.i.d.\ RV's, without
any structure in the form of (\ref{ham}) or its above--described extensions. In particular, 
in the absence of a magnetic field,
the $2^n$ RV's $\{\calE(\bx)\}$ are taken to be zero--mean Gaussian RV's, all with variance
$nJ^2/2$, where $J$ is a parameter.\footnote{The variance scales linearly with $n$ to match
the behavior of the Hamiltonian (\ref{ham}) with 
a limited number of interacting neighbors and random interaction parameters, which has a number of
independent terms that is linear in $n$.} The beauty of the REM is in that on the one hand, it
is very easy to analyze, and on the other hand, it consists of sufficient richness to exhibit
phase transitions. 

The basic observation about the REM is that for a typical realization of the
configurational energies $\{\calE(\bx)\}$, the number of configurations with energy 
about $E$ (i.e., between $E$ and $E+dE$), $N(E)$, is proportional 
(up to sub--exponential terms in $n$) to $2^n\cdot e^{-E^2/(nJ^2)}$, as long as
$|E|\le E_0\dfn nJ\sqrt{\ln 2}$, whereas energy levels outside this range are typically
not populated by spin configurations ($N(E)=0$), as the probability 
of having at least one configuration with such an energy decays exponentially with $n$.
Thus, the asymptotic (thermodynamical) entropy per spin, which is defined by
$$S(E)=\lim_{n\to\infty}\frac{\ln N(E)}{n}$$
is given by
$$S(E)=\left\{\begin{array}{ll}
\ln 2 -\left(\frac{E}{nJ}\right)^2 & |E|< E_0\\
0 & |E|= E_0\\
-\infty & |E| > E_0 \end{array}\right.$$
The partition function of a typical realization of a REM spin glass
is then
\begin{eqnarray}
Z(\beta)&\exe&\int_{-E_0}^{E_0}dE\cdot N(E)\cdot e^{-\beta E}\nonumber\\
&\exe&\int_{-E_0}^{E_0}dE \cdot e^{nS(E)}\cdot e^{-\beta E}
\end{eqnarray}
whose exponential growth rate,
$$\phi(\beta)\dfn\lim_{n\to\infty}\frac{\ln Z(\beta)}{n},$$
behaves according to 
\begin{eqnarray}
\phi(\beta)
&=&\max_{|E|\le E_0}\left[S(E)-\beta\cdot\frac{E}{n}\right]\nonumber\\
&=&\max_{|E|\le E_0}\left[\ln 2-
\left(\frac{E}{nJ}\right)^2-\beta J \cdot\left(\frac{E}{nJ}\right)\right].
\end{eqnarray}
Solving this simple optimization problem, we find that $\phi(\beta)$
is given by
$$\phi(\beta)=\left\{\begin{array}{ll}
\ln 2 +\frac{\beta^2J^2}{4} & \beta \le \frac{2}{J}\sqrt{\ln 2}\\
\beta J\sqrt{\ln 2} & \beta > \frac{2}{J}\sqrt{\ln 2}\end{array}\right.$$
which means that the asymptotic free energy per spin, a.k.a.\ the {\it free energy density},
which is obtained by
$$F(\beta)=-\frac{\phi(\beta)}{\beta},$$
is given by (cf.\ \cite[Proposition 5.2]{MM06}):
$$F(\beta)=
\left\{\begin{array}{ll}
-\frac{\ln 2}{\beta}-\frac{\beta J^2}{4} & \beta \le \frac{2}{J}\sqrt{\ln 2}\\
-J\sqrt{\ln 2} & \beta > \frac{2}{J}\sqrt{\ln 2}\end{array}\right.$$
Thus, the free energy density is subjected to a phase transition at the inverse
temperature $\beta_0\dfn \frac{2}{J}\sqrt{\ln 2}$. At high temperatures ($\beta < \beta_0$),
which is referred to as the {\it paramagnetic phase}, the partition function is dominated by 
an exponential number of configurations with energy $E=-n\beta J^2/2$ and the entropy
grows linearly with $n$. When the system is cooled to $\beta=\beta_0$ and beyond, 
which is the {\it glassy phase},
the system freezes but it is still in disorder -- the partition function
is dominated by a subexponential number of configurations of minimum energy $E=-E_0$.
The entropy, in this case, grows sublinearly with $n$, namely the entropy 
per spin vanishes, and the free energy density no longer depends on $\beta$.
Further details about the REM can be found in \cite{MM06} and the references mentioned in
the Introduction.

\subsection{The REM and Random Code Ensembles}
\label{remc}

As described in \cite{MM06}, there is an interesting analogy between the REM
and the partition function pertaining to {\it finite temperature decoding} \cite{Rujan93}
of ensembles of channel block codes (see also \cite{Merhav07}).

In particular, consider a codebook $\calC$ of $M=e^{nR}$ binary codewords 
of length $n$, $\bx_1,\ldots,\bx_M$,
to be used across a binary symmetric channel (BSC)
with crossover probability $p$. Given a binary vector $\by$ at the channel output,
consider the generalized posterior parametrized by $\beta$: 
\begin{eqnarray}
\label{pbeta}
P_\beta(\bx|\by)&=&\frac{P^\beta(\by|\bx)}{\sum_{\bx'\in\calC}
P^\beta(\by|\bx')}\nonumber\\
&=&\frac{e^{-\beta Bd_H(\bx,\by)}}{\sum_{\bx'\in\calC}
e^{-\beta Bd_H(\bx',\by)}}\nonumber\\
&\dfn&\frac{e^{-\beta B d_H(\bx,\by)}}{Z(\beta|\by)},
\end{eqnarray}
where $B\dfn\ln\frac{1-p}{p}$, $d_H(\bx,\by)$ is the Hamming distance between $\bx$ and $\by$, and
where the real posterior is obtained, of course, for $\beta=1$. This is identified as a Boltzmann
distribution whose energy function (which depends on the given $\by$) is
$\calE(\bx)=Bd_H(\bx,\by)$.
As described in \cite{MM06} and \cite{Merhav07},
there are a few motivations for introducing the temperature parameter $\beta$ here.
First, it allows a degree of freedom
in case there is some uncertainty regarding the channel
noise level (small $\beta$ corresponds to high noise level). Second, it is
inspired by the ideas behind simulated annealing techniques: by
sampling from $P_\beta$ while gradually increasing $\beta$
(cooling the system), the minima of the energy function
(ground states) can be found.
Third, by applying symbolwise MAP decoding, i.e., decoding the $\ell$--th
symbol of $\bx$ as
$\mbox{arg}\max_a P_\beta(x_\ell=a|\by)$, where
$$P_\beta(x_\ell=a|\by)=
\sum_{\bx\in\calC:~x_\ell=a}P_\beta(\bx|\by),$$
we obtain
a family of
{\it finite--temperature decoders}
parametrized by $\beta$, where $\beta=1$ corresponds
to minimum symbol error probability (with respect to the true channel)
and $\beta\to\infty$ corresponds to minimum block error probability.
As in \cite{MM06}, we will distinguish between two contributions of $Z(\beta|\by)$:
One is $Z_c(\beta|\by)=e^{-\beta Bd_H(\bx_0,\by)}$, where $\bx_0$ is the actual codeword
transmitted, and the other is $Z_e(\beta|\by)=\sum_{\bx'\in\calC\setminus{\bx_0}}e^{-\beta Bd_H(\bx',\by)}$,
pertaining to all incorrect codewords. The former is typically about $e^{-\beta Bnp}$
since $d_H(\bx_0,\by)$ concentrates about $np$. We next focus on the behavior of $Z_e(\beta|\by)$.

To this end, consider a random selection of the code $\calC$, where every bit of every codeword
is drawn by an independent fair coin tossing. 
For a given $\by$, the energy levels $\{Bd_H(\bx,\by)\}$
pertaining to all incorrect codewords are RV's 
(exactly like in the REM) because of the random selection of these
codewords. Now, the total number of correct codewords is about $e^{nR}$,
and the probability that a randomly chosen $\bx$ would fall at distance $d=n\delta$ from $\by$
is exponentially $e^{n[h(\delta)-\ln 2]}$, where
$$h(\delta)=-\delta\ln \delta-(1-\delta)\ln(1-\delta),$$
then the typical number of codewords at normalized distance $\delta$ is about
$$N(\delta)=e^{n[R+h(\delta)-\ln 2]}$$
as long as $R+h(\delta)-\ln 2 \ge 0$ and $N(\delta)=0$ when $R+h(\delta)-\ln 2<0$.
Thus, letting $\delta(R)$ denote the small solution to the equation $R+h(\delta)-\ln 2=0$
(the Gilbert--Varshamov distance),
we find that, with a clear analogy to the REM,
the corresponding thermodynamical entropy is given by
\begin{equation}
\label{srem}
S(\delta)=\left\{\begin{array}{ll}
R+h(\delta)-\ln 2 & \delta(R)<\delta < 1-\delta(R)\\
0 & \delta=\delta(R)~~\mbox{or}~~\delta = 1-\delta(R)\\
-\infty & \delta<\delta(R)~~\mbox{or}~~\delta > 1-\delta(R)
\end{array}\right.
\end{equation}
Accordingly, the partition function $Z_e(\beta|\by)$ of a typical code is given by
\begin{equation}
\label{ze}
Z_e(\beta|\by)\exe \sum_{\delta=\delta(R)}^{1-\delta(R)} e^{n[R+h(\delta)-\ln 2]}\cdot
e^{-\beta Bn\delta}\exe \exp\{n[R-\ln 2+\max_{\delta(R)\le\delta\le 1-\delta(R)}(h(\delta)-
\beta B\delta)]\},
\end{equation}
and the free energy density pertaining to $Z_e$ behaves according to 
\begin{equation}
F_e(\beta)= \left\{\begin{array}{ll}
\frac{\ln 2-R-h(p_\beta)}{\beta}+Bp_\beta & \beta \le \beta_0\\
B\delta(R) & \beta > \beta_0 \end{array}\right.
\end{equation}
where 
$$p_\beta=\frac{p^\beta}{p^\beta+(1-p)^\beta}$$
and
$$\beta_0=\frac{\ln[(1-\delta(R))/\delta(R)]}{B},$$
and where, again, the first line of $F_e(\beta)$ corresponds to the paramagnetic phase
with exponentially many codewords at distance (energy) $np_\beta$ from $\by$,
and the second line is the glassy phase with subexponentially many codewords at distance
$n\delta(R)$. In \cite{Merhav07}, these free energies
are related to random coding exponents as mentioned in the Introduction.

By the same token, in rate--distortion source coding, if one defines 
the partition function as
$$Z(\beta)=\sum_{\hat{\bx}\in\calC} e^{-\beta d_H(\bx,\hat{\bx})}$$
with $\bx$ being the source vector, $\{\hat{\bx}\}$ being the reproduction codevectors,
and $d_H(\bx,\by)$ being the Hamming distortion measure, then the same analysis takes place.
In the sequel, we will motivate this definition of the partition function 
of rate--distortion coding and use it.

\subsection{The GREM}
\label{grem}

As we have seen, the REM is an extremely simple model to analyze, but its
simplicity is also recognized as a drawback from the aspect of faithfully modeling a spin glass.
The reason for this is the lack of structure which is needed to allow dependencies between
energy levels of spin configurations that are closely related: For example, if $\bx$ and $\bx'$ differ
only in a single component, it is conceivable that the respective energies would be
close, as suggested by (\ref{ham}). To this end, as described in the Introduction, Derrida
proposed a generalized version of the REM -- the GREM, which introduces dependencies between
configurational energies in an hierarchical fashion. We next briefly review the GREM.

A GREM with $k$ levels can best be thought of as a tree 
with $2^n$ leaves and depth $k$, where
each leaf represents one spin configuration. This tree 
is defined by $k$ positive parameters, $\alpha_1,\ldots,\alpha_k$,
which are all in the interval $(1,2)$, and whose product, $\prod_{i=1}^k\alpha_i$, equals $2$.
The construction of this tree is as follows:
The root of the tree is connected to $\alpha_1^n$ distinct
nodes,\footnote{We are approximating $\alpha_1^n, \alpha_2^n,\ldots \alpha_k^n$ by integers.}
which will be referred to as first--level nodes. Each first--level
node is in turn connected to $\alpha_2^n$
distinct second--level nodes, thus a total of
$(\alpha_1\alpha_2)^n$ second--level nodes. In the case $k=2$, these second--level nodes are
the leaves of the tree and $\alpha_1\alpha_2=2$. If $k > 2$, the process continues, and
each second--level node is connected to $\alpha_3^n$ third--level nodes, and so on. At the
last step, each one of the $\prod_{i=1}^{k-1}\alpha_i^n$ nodes at level $k-1$ is connected
to $\alpha_k^n$ distinct leaves, thus a total of $\prod_{i=1}^k\alpha_i^n=2^n$ leaves. 
The REM corresponds 
to the degenerate special
case where $k=1$. 

The random selection of energy levels for the GREM is defined by another
set of $k$ parameters, $a_1,a_2,\ldots,a_k$, which are all positive reals that sum to unity.
The random selection is carried out in the following manner:
For each one of the $\prod_{j=1}^i\alpha_j^n$ branches 
emanating from $(i-1)$--th level nodes and connecting them to $i$--th level nodes ($i=1,2,\ldots,k$)
in the tree, we randomly choose an 
independent RV, henceforth referred to as a {\it branch energy},
which is a zero mean, Gaussian RV with variance $nJ^2a_i/2$, 
where $J$ is like in the REM and where $\{a_i\}_{i=1}^k$ are as described above.
Finally, the energy level of a given configuration 
is given by the sum of branch energies along the path from the root to the leaf that
represents this configuration. Thus, the total energy, is the sum of $k$ independent
zero--mean Gaussian RV's with variances $nJ^2a_i/2$, and so, it is 
zero--mean Gaussian RV with variance $nJ^2/2$, exactly like in the REM.
However, now the energy levels of different configurations may be clearly correlated if the
paths from the root to their corresponding leaves share some common branches before they split.
The degree of statistical dependence is according to their distance along the tree. For example,
if two configurations are first--degree siblings, i.e., they share the same parent node at level
$k-1$, then all their energy components are the same except 
their last branch energies, which are independent. On the other extreme, if their paths are
completely distinct, then their energies are independent.

The GREM for $k=2$ is analyzed in \cite{Derrida85}. 
We next present the derivation for this case (with a few more details than in \cite{Derrida85}).
Let $\alpha_1$ and $\alpha_2$ be positive numbers whose product equals $2$,
and let $a_1$ and $a_2$ be positive numbers whose sum equals $1$.
Now, every configuration with energy $E$ has some first--level branch energy $\epsilon$
and second--level branch energy $E-\epsilon$. For a typical realization of this GREM,
the number of first--level branches with energy about $\epsilon$ is exponentially
$$N_1(\epsilon)\exe\alpha_1^n\cdot\exp\left\{-\frac{\epsilon^2}{nJ^2a_1}\right\}=
\exp\left\{n\left[\ln\alpha_1-\frac{1}{a_1}\left(\frac{\epsilon}{nJ}\right)^2\right]\right\},$$
provided that the expression in the square brackets is non--negative, i.e.,
$|\epsilon|\le \epsilon_0\dfn nJ\sqrt{a_1\ln\alpha_1}$, and $N_1(\epsilon)=0$ otherwise. 
Therefore, the number of configurations with total energy about $E$
is exponentially
$$N_2(E)\exe\int_{-\epsilon_0}^{\epsilon_0} d\epsilon\cdot N_1(\epsilon)\cdot
\exp\left\{n\left[\ln\alpha_2-\frac{1}{a_2}\left(\frac{E-\epsilon}{nJ}\right)^2\right]\right\},$$
whose exponential rate (the entropy per spin) is given by
$$S(E)=\lim_{n\to\infty}\frac{\ln N_2(E)}{n}=
\max_{|\epsilon|\le\epsilon_0}\left[\ln\alpha_1-\frac{1}{a_1}\left(\frac{\epsilon}{nJ}\right)^2+
\ln\alpha_2-\frac{1}{a_2}\left(\frac{E-\epsilon}{nJ}\right)^2\right].$$
Note that $S(E)$
is an even function, non--increasing in $|E|$, and it should be
kept in mind that beyond the value of $|E|$ at which $S(E)$ vanishes, denote it by $\hat{E}$,
we have $S(E)=-\infty$ since $N_2(E)$ is typically zero (as was the case with the REM). 
We shall get back to this point
shortly, but for a moment, let us ignore it and solve
the maximization problem pertaining to the above expression of $S(E)$, as is.
Denoting the resulting maximum by $\tilde{S}(E)$ (to distinguish from $S(E)$, where $\hat{E}$
and the jump to $\infty$ are taken into account), we get:
\begin{equation}
\label{psibasic}
\tilde{S}(E)=\left\{\begin{array}{ll}
\ln 2 -\left(\frac{E}{nJ}\right)^2 & |E|\le E_1\\
\ln\alpha_2-\frac{1}{a_2}\left(\frac{E}{nJ}-\sqrt{a_1\ln\alpha_1}\right)^2 & |E|> E_1
\end{array}\right.
\end{equation}
where $E_1\dfn nJ\sqrt{(\ln\alpha_1)/a_1}$. Taking now into account the above mentioned
observation concerning the criticality of the point $|E|=\hat{E}$, we have to distinguish between
two cases. The first is the case where $\hat{E} < E_1$, namely, the first line of the
above expression of $\tilde{S}(E)$ vanishes for $|E|$ smaller than $E_1$. The first line
vanishes for $|E|=E_0=nJ\sqrt{\ln 2}$, so the condition for this case to hold is 
$E_0\le E_1$, or equivalently, $(\ln\alpha_1)/a_1 \ge \ln 2$. In this case,
we then have:
$$S(E)=\left\{\begin{array}{ll}
\ln 2 -\left(\frac{E}{nJ}\right)^2 & |E|\le E_0\\
0 & |E|=E_0\\
-\infty & |E| > E_0 \end{array}\right.$$
which is exactly the same behavior as in the ordinary REM ($k=1$).
Consequently, the exponential rate of the partition function, which is given by
$$\phi(\beta)=\lim_{n\to\infty}\frac{\ln Z(\beta)}{n}=\max_E\left[S(E)-\beta\frac{E}{n}\right],$$
is also the same as in the REM, namely,
$$\phi(\beta)=\left\{\begin{array}{ll}
\ln 2+\frac{\beta^2J^2}{4} & \beta < \beta_0\\
\beta J\sqrt{\ln 2} & \beta\ge\beta_0\end{array}\right.$$
where $\beta_0$ is the above defined critical inverse temperature of the REM (see Subsection \ref{rem}).

We next consider the complementary case where $(\ln\alpha_1)/a_1 < \ln 2$.
In this case, the expression of $S(E)$ should take into account
the fact that it vanishes (and then becomes $-\infty$) according to the
second line of (\ref{psibasic}). This amounts to:
\begin{equation}
\label{2phasetransitions}
S(E)=\left\{\begin{array}{ll}
\ln 2 -\left(\frac{E}{nJ}\right)^2 & |E|\le E_1\\
\ln\alpha_2-\frac{1}{a_2}\left(\frac{E}{nJ}-\sqrt{a_1\ln\alpha_1}\right)^2 & E_1\le |E|< E_2\\
0 & |E|= E_2\\
-\infty & |E| > E_2
\end{array}\right.
\end{equation}
where $E_2\dfn nJ(\sqrt{a_1\ln\alpha_1}+\sqrt{a_2\ln\alpha_2})$.
Before we compute the corresponding partition function,
we make the following observation:
$$\ln 2=\frac{\ln\alpha_1+\ln\alpha_2}{a_1+a_2}\le \max_{i=1,2}\frac{\ln\alpha_i}{a_i},$$
where the inequality follows from the well--known inequality
$[\sum_{i=1}^m a_i]/[\sum_{i=1}^m b_i]\le\max_{1\le i \le m}a_i/b_i$ for positive $\{a_i\}$ and
$\{b_i\}$ \cite[Lemma 1]{CO96}. In the same manner, using the similar inequality
$[\sum_{i=1}^m a_i]/[\sum_{i=1}^m b_i]\ge\min_{1\le i \le m}a_i/b_i$, we get
$$\ln 2\ge \min_{i=1,2}\frac{\ln\alpha_i}{a_i}.$$
It follows then that the condition $(\ln\alpha_1)/a_1 < \ln 2$ is equivalent
to the condition $(\ln\alpha_1)/a_1 < \ln 2 < (\ln\alpha_2)/a_2$.
Defining
$$\beta_i=\frac{2}{J}\sqrt{\frac{\ln \alpha_i}{a_i}},~~i=1,2$$
we then have $\beta_1 < \beta_0 < \beta_2$.
Let us examine how $\phi(\beta)$
behaves as $\beta$ grows from zero to infinity. 
For small enough $\beta$, the achiever of $\phi(\beta)$,
call it $E^*$, is still smaller in absolute value than $E_0$, and then it is
obtained from equating to zero the derivative of $[S(E)-\beta E/n]$, with $S(E)$
being according to first line of (\ref{2phasetransitions}), thus $E^*=-\frac{n}{2}\beta J^2$.
This remains true as long as $\frac{n}{2}\beta J^2\le E_1$, which means $\beta\le \beta_1$.
In this case, the partition function is dominated by 
$\exp\{n[\ln\alpha_1-a_1\beta^2J^2/4]\}$ first--level branches with energy
$\epsilon^*=-\frac{a_1}{2}n\beta J^2$, each followed by
$\exp\{n[\ln\alpha_1-a_1\beta^2J^2/4]\}$ second--level branches with energy
$E^*-\epsilon^*=-\frac{a_2}{2}n\beta J^2$, and this is a pure paramagnetic phase.
As $\beta$ continues to grow beyond $\beta_1$, but is still below $\beta_2$, the partition
function is dominated by a subexponential number of first--level branches of energy
$-nJ\sqrt{a_1\ln\alpha_1}$ followed by $\exp\{n[\ln\alpha_1-a_1\beta^2J^2/4]\}$
second--level branches with energy $E^*+nJ\sqrt{a_1\ln\alpha_1}$. This is a ``semi--glassy''
phase, where the first--level branches are already glassy but the second--level ones are
still paramagnetic.
As $\beta$ exceeds $\beta_2$, this becomes a pure glassy phase where the partition function
is dominated by a subexponential number of first--level branches with energy
$-nJ\sqrt{a_1\ln\alpha_1}$ and
a subexponential number of second--level branches with energy
$-nJ\sqrt{a_2\ln\alpha_2}$. Accordingly, the function $\phi(\beta)$ exhibits two phase transitions
at inverse temperatures $\beta_1$ and $\beta_2$:
$$\phi(\beta)=\left\{\begin{array}{ll}
\ln 2+\frac{\beta^2J^2}{4} & \beta < \beta_1\\
\beta J\sqrt{a_1\ln\alpha_1}+\ln\alpha_2+\frac{a_2\beta^2J^2}{4} & \beta_1\le\beta<\beta_2\\
\beta J(\sqrt{a_1\ln\alpha_1}+\sqrt{a_2\ln\alpha_2}) & \beta\ge \beta_2 \end{array}\right.$$
Again, the free energy density is obtained by $F(\beta)=-\phi(\beta)/\beta$.

This different behavior of the GREM for the two different cases will
be pivotal to our later discussion on the parallel behavior of ensembles of codes.
When there is a general number $k$ of levels, the above analysis of the GREM
becomes, of course, more complicated and there are more cases to consider, but the
concepts remain the same. There can be up to $k$ phase transitions, but there can
be less, depending on the parameters of the model $\{a_i,\alpha_i\}_{i=1}^k$.
For details, the reader is referred to \cite{DG86a},\cite{DG86b}.

\section{Relations Between GREM and Hierarchical Code Ensembles}
\label{main}

In analogy to the relationship between the REM and ordinary 
ensembles of block codes, as was described
in Subsection \ref{remc}, it is natural to wonder about the possibility of similar
relationships between the GREM and more general ensembles of block codes,
and to ask whether the fact that the GREM exhibits
different types of behavior (as we have seen in Subsection \ref{grem}), has
implications on the behavior of these ensembles of codes. 
Since the GREM is defined by an hierarchical (tree) structure, it is plausible
to expect that if a relationship to coding exists, it will be in the context of
ensembles of codes which have hierarchical structures as well. 
Hierarchically structured ensembles of codes are encountered in numerous applications
in Information Theory, including block--causal tree--structured 
source codes and channel codes of the type described informally 
in the Introduction, successive refinement
source codes \cite{EC91},\cite{Koshelev80},\cite{Rimoldi94},
codes for the broadcast channel \cite[Chap.\ 15.6]{CT06} and codes based on binning techniques
(see, e.g., \cite{WZ76},\cite{GP80},\cite{Wyner75}),
just to name a few. In this paper, we confine our attention to the first above--mentioned class of codes.

The fact that the GREM
behaves, in some situations, like the REM, and the REM is analogous to an ordinary
block code without any hierarchical structure (cf.\ \ref{remc}), may hint that
in the parallel situations in the realm of our coding problem, a typical code
from the hierarchical ensemble will perform essentially as well as a typical (good) code
without the hierarchical structure. In these situations then (which can be imposed by
a clever choice of certain design parameters), 
it would be interesting to explore the question whether we may enjoy the benefit that
the hierarchical structure buys us (in our case, reduced delay) without 
essentially paying in terms of performance. As we show in this section, the answer
to this question turns out to be affirmative to a large extent, both in the source coding
setting and in the channel coding setting.

Finally, in closing this introductory part of Section \ref{main}, a more technical comment is in order:
As in Subsection \ref{remc}, throughout the sequel, 
we confine ourselves to the memoryless binary symmetric source (BSS)
with the Hamming distortion measure, in the context of source coding,
and to the binary symmetric channel (BSC) in the context of channel coding. 
The random coding distribution in both problems will be i.i.d.\ and uniform,
i.e., each bit of each codeword will be drawn by independent fair coin tossing.
Also, we will focus mostly on the case $k=2$.
The reason for this is that
our purpose is this paper
is more to demonstrate certain concepts, and so,
we prefer to slightly sacrifice generality at the benefit of simplicity, 
and so, better readability, and a smaller amount of space. Having said that,
all the derivations can be extended
to apply to more general memoryless sources, channels,
and random coding distributions (as was done in \cite{Merhav07}), as well
as to a general number $k$ of stages.

\subsection{Lossy Source Coding}
\label{lossysourcecoding}

Consider the BSS $X_1,X_2,\ldots$, $X_i\in\{0,1\}$ ($i$ -- positive integer) 
and the Hamming distortion measure between two binary $n$--vectors $\bx$ and $\hat{\bx}$:
$$d_H(\bx,\hat{\bx})=\sum_{i=1}^n d_H(x_i,\hat{x}_i),$$ 
where $d_H(a,b)=1$ if $a\ne b$ and $d_H(a,b)=0$ if $a=b$, $a,b\in\{0,1\}$.
Before discussing ensembles of codes with hierarchical structures, let
us first confine attention to an ordinary ensemble with no structure.

Consider a random selection of a codebook of size $M=e^{nR}$ ($R$ being 
the coding rate in nats per source bit), $\calC=\{\hbx_1,\ldots,\hbx_M\}$,
$\hbx_i\in\{0,1\}^n$, $i=1,2,\ldots,M$,
where each component of each codeword is drawn randomly by an independent
fair coin tossing. For a given source vector $\bx$ and for a given such
randomly drawn codebook $\calC$, let $\Delta(\bx)=\min_{\hbx\in\calC}d_H(\bx,\hbx)$
denote the distortion associated with encoding $\bx$.

Instead of examining the expected distortion, $\bE\{\Delta(\bX)\}$, w.r.t.\ both
the source and the random codebook selection, as is traditionally done,
we will concern ourselves with a more refined and more informative objective function,
which is the characteristic function of $\Delta(\bX)$, namely,
$$\Psi_n(s,R)=\bE\{\exp[-s\Delta(\bX)]\},$$ 
or in particular, its exponential rate
$$\psi(s,R)=-\lim_{n\to\infty}\frac{\ln\Psi_n(s,R)}{n}$$
focusing on the range $s\ge 0$. As is well known, the characteristic function provides information
not only on the expected distortion, $\bE\{\Delta(\bX)\}$, but also on every moment of $\Delta(\bX)$
(by taking derivatives of $\Psi_n(s,R)$ at $s=0$). It is also intimately related to the tail
behavior (i.e., large deviations probabilities) of the distribution of $\Delta(\bX)$ via Chernoff
bounds.

In order to analyze $\Psi_n(s,R)$ and then $\psi(s,R)$, first, for an ordinary ensemble, and later for
an hierarchical structured ensemble, it is convenient to define, 
for given $\bx$ and $\calC$, the partition function\footnote{For a given $\bx$, the
partition function $Z(\beta|\bx)$ induced by a typical codebook is exactly the same as in
(\ref{ze}), with the minor modification that here $\beta$ is not scaled by $B$ as in (\ref{ze}).}
\begin{equation}
Z(\beta|\bx)=\sum_{\hbx\in\calC} e^{-\beta d_H(\bx,\hbx)}.
\end{equation}
The function $\Psi_n(s,R)$ is obtained from the partition function by
$$\Psi_n(s,R)=
\bE\{\lim_{\theta\to\infty}Z^{1/\theta}(s\cdot\theta|\bX)\}=
\lim_{\theta\to\infty}\bE\{Z^{1/\theta}(s\cdot\theta|\bX)\}.$$
In the definition of the ensemble behavior of $\psi(s,R)$, there are now two options. The first
is to think of the above defined expectation of $Z^{1/\theta}(s\theta|\bX)$ as being
taken w.r.t.\ both the source $\bX$ and the code ensemble $\{\calC\}$, and then to define
$\psi(s,R)$ as above. The second option is to define the above expectation of 
$Z^{1/\theta}(s\theta|\bX)$ w.r.t.\ the source only, while keeping $\calC$ fixed, and
then to define $\psi(s,R)$ as $-\lim_{n\to\infty}\bE\{\ln\Psi_n(s,R)\}/n$, where the latter expectation
is across the ensemble of codebooks $\{\calC\}$. The difference between meanings of the two approaches
is in the point of view: In the former approach the randomness of both $\bX$ and $\calC$
are treated on equal grounds, and this makes sense if $\bX$ and $\calC$ vary on the same
time scale (e.g., when the codebook varies frequently according to
some secret key). In the parallel discussion on spin glasses (cf.\ Section 3.1),
this is analogous to the double randomness of both the spin configuration and the interaction
parameters, and in the language of statistical physicists, this is called {\it annealed}
averaging. The second approach, which physicists refer to as {\it quenched} averaging, fits
better the paradigm where the code $\calC$ is held fixed over many realizations of the source $\bX$.
In the Information Theory literature, it is more customary to adopt an approach analogous to
annealed averaging\footnote{In particular, source and channel random coding exponents are
normally defined as exponential rates of ensemble--average error 
probabilities, and not as ensemble--average
exponents of error probabilities.} and so, we shall do the same here.

\subsubsection{The Ordinary Ensemble}
\label{ordinary}

\begin{figure}[ht]
\hspace*{4cm}\input{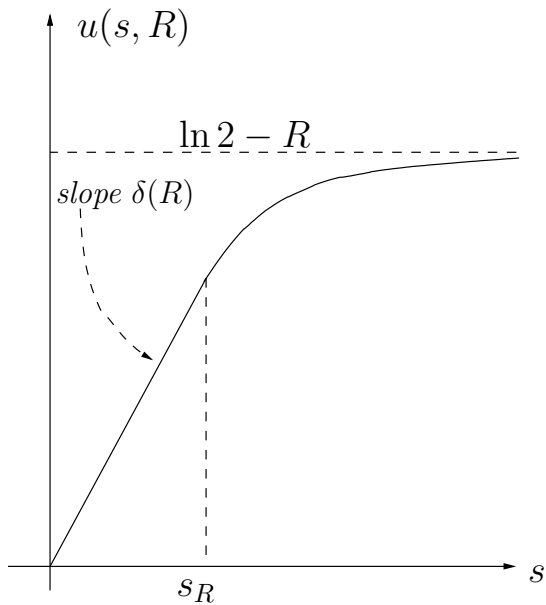}
\caption{$u(s,R)$ as a function of $s$ for fixed $R$.}
\label{gen}
\end{figure}

Let us begin the with the calculation of the annealed version of $\psi(s,R)$, first, for a
an ordinary non--hierarchical code:
\begin{eqnarray}
\bE\{Z^{1/\theta}(s\theta|\bX)\}&=&
\bE\left\{\left[\sum_{\hbx\in\calC}\exp(-s\theta d_H(\bX,\hbx))\right]^{1/\theta}\right\}\nonumber\\
&=&\bE\left\{\left[\sum_{d=0}^nN(d)\cdot e^{-s\theta d}\right]^{1/\theta}\right\}\nonumber\\
&\exe&\bE\left\{\sum_{d=0}^nN^{1/\theta}(d)\cdot e^{-sd}\right\}\nonumber\\
&=&\sum_{d=0}^n\bE\{N^{1/\theta}(d)\}\cdot e^{-sd}
\end{eqnarray}
where $N(\delta)$ is the number of codewords whose normalized Hamming distance from $\bX$
is exactly $\delta$, and
where the third (exponential) equality
holds, even before taking the expectation,
because the summation over $d$ consists of a {\it subexponential} number of terms, and so, both
$[\sum_d N(d)e^{-s\theta d}]^{1/\theta}$ and
$\sum_d N^{1/\theta}(d)e^{-sd}$ are of the same exponential order as
$\max_d N^{1/\theta}(d)e^{-sd}=
[\max_d N(d)e^{-s\theta d}]^{1/\theta}$. This is different from the original summation over $\calC$
which contains an {\it exponential} number of terms.
Now, as is shown in Subsection A.1 of the Appendix (see also \cite{Merhav07b}),
\begin{equation}
\label{moments}
\bE\{N^{1/\theta}(n\delta)\}\exe
\left\{\begin{array}{ll}
e^{n[R+h(\delta)-\ln 2]} & \delta < \delta(R)~~\mbox{or}~~ \delta > 1-\delta(R)\\
e^{n[R+h(\delta)-\ln 2]/\theta} & \delta(R)\le\delta\le 1-\delta(R) 
\end{array}\right.
\end{equation}
where $\delta(R)$ is defined (cf.\ Subsection \ref{remc}) as the small solution to
the equation $R+h(\delta)-\ln 2=0$, which is also the distortion--rate function
of the BSS. This gives
\begin{eqnarray}
\bE\{Z^{1/\theta}(s\theta|\bX)\}&\exe&
\sum_{\delta<\delta(R)} e^{n[R+h(\delta)-\ln 2]}\cdot e^{-s\delta n}+
\sum_{\delta\ge\delta(R)} e^{n[R+h(\delta)-\ln 2]/\theta}\cdot e^{-s\delta n}\nonumber\\
&\dfn&A+B
\end{eqnarray}
Now, as $\theta\to\infty$, the term $B$ tends to $\sum_{\delta\ge\delta(R)}e^{-s\delta n}$,
which is of the exponential order of $e^{-ns\delta(R)}$. The term $A$, which is independent
of $\theta$, is of the exponential order of $e^{-nu(s,R)}$, where
$$u(s,R)\dfn\ln 2-R-\max_{\delta\le\delta(R)}[h(\delta)-s\delta]=
\left\{\begin{array}{ll}
s\delta(R) & s\le s_R\\
v(s,R) & s > s_R \end{array}\right.$$
where
$$s_R\dfn\ln\left[\frac{1-\delta(R)}{\delta(R)}\right].$$
and 
$$v(s,R)\dfn \ln 2-R+s-\ln(1+e^s).$$
Since $v(s,R)$ never exceeds $s\delta(R)$ for $s> s_R$, the dominant term is $A$,
and therefore, for the ordinary block code ensemble, we have:
$$\psi(s,R)=u(s,R).$$ 
It is not difficult to show also, using sphere covering considerations,
that $u(s,R)$ is the best achievable performance in terms
of the exponential rate of the characteristic function of the distortion.
The function $u(s,R)$ is depicted qualitatively in Fig.\ \ref{gen}.

\subsubsection{The Hierarchical Ensemble}
\label{hierarchical}

We proceed to define the ensemble of hierarchical codes and to analyze
its performance with relation to the GREM. Let $n=n_1+n_2$, where $n$, $n_1$
and $n_2$ are positive integers. For a given $R_1$,
consider a random selection of a codebook of size $M_1=e^{n_1R_1}$,
$\calC_1=\{\hbx_1,\ldots,\hbx_{M_1}\}$,
$\hbx_i\in\{0,1\}^{n_1}$, $i=1,2,\ldots,M_1$,
where each component of each codeword is drawn randomly by an independent
fair coin tossing. Next, given $R_2$, for each $i=1,2,\ldots,M_1$,
consider a similar random selection of a codebook of size $M_2=e^{n_2R_2}$,
$\calC_2(i)=\{\tbx_{i,1},\ldots,\tbx_{i,M_2}\}$,
$\tbx_{i,j}\in\{0,1\}^{n_2}$, $j=1,2,\ldots,M_2$.

The encoder works as follows: Given a source vector $\bx\in\{0,1\}^n$,
it finds a pair of indices $(i,j)$, $i=1,2,\ldots,M_1$, $j=1,2,\ldots,M_2$,
such that the distortion between $\bx$ and the concatenation of the codewords
$(\hbx_i,\tbx_{i,j})$ is minimum. The index $i$ is encoded by $n_1R_1$ nats
and the index $j$ (given $i$) is encoded by $n_2R_2$ nats, thus a total of
$nR\dfn n_1R_1+n_2R_2$ nats, where $R$ is the overall rate, given by 
$$R=\lambda R_1+(1-\lambda)R_2, ~~~\lambda=\frac{n_1}{n}.$$
The decoder can, of course, generate the first--stage reproduction
$\hbx_i$ based on the first $n_1R_1$ nats received,
without having to wait for the $n_2R_2$ following ones. The extension of this hierarchical
structure to a larger number of stages $k$ should be obvious. In particular, 
as mentioned in the Introduction, if $k$ divides $n$ 
and the $n$--block is divided to $k$ sub--blocks of length $n/k$ each, then the decoder
can generate chunks of the reproduction at a reduced delay of $n/k$ instead of $n$.

The analogy of this structure with the GREM should also be obvious. The code has a tree
structure and the configurational energies of the GREM play the same role as the distortion
here, as the overall distortion is the cumulative sum of the per--stage distortions.
Also, the coding rate $R_i$ here plays the same role as $\ln\alpha_i$ of the GREM ($i=1,2$).
Thus, it is natural to expect that the partition function $Z(\beta|\bx)$ of this code ensemble would 
behave analogously to that of the GREM, as we shall see next.

For the sake of simplicity, we return to the case $k=2$, with the understanding that
our derivations can be extended without any essential difficulties to a general $k$.
Before analyzing the characteristic function of the distortion along with its exponential rate,
it is instructive to examine the partition function $Z(\beta|\bx)$ for a given $\bx$ and address
the analogy with that of the GREM.

For a given $\bx$ and a typical code in the ensemble, 
there are $N_1(\delta_1)\exe e^{n_1[R_1+h(\delta_1)-\ln 2]}$
first-stage codewords $\{\hbx\}$ at distance $n_1\delta_1$ from 
the vector formed by the first $n_1$ components of $\bx$, provided that $\delta_1\ge \delta(R_1)$
and $N_1(\delta_1)=0$ otherwise. For each one of these first--stage codewords, there are
$e^{n_2[R_2+h(\delta_2)-\ln 2]}$ second--stage codewords $\{\tbx\}$ at distance $n_2\delta_2$ from
the vector formed by the last $n_2$ components of $\bx$, provided that $\delta_2\ge \delta(R_2)$.
Thus, the total number of concatenated codewords $\{(\hbx,\tbx)\}$ at distance $n\delta=n_1\delta_1+n_2
\delta_2$ (that is, $\delta=\lambda\delta_1+(1-\lambda)\delta_2$) from $\bx$ is given by
\begin{eqnarray}
N_2(\delta)&\exe&\sum_{\delta_1=\delta(R_1)}^{1-\delta(R_1)} e^{n_1[R_1+h(\delta_1)-\ln 2]}\cdot
e^{n_2[R_2+h((\delta-\lambda\delta_1)/(1-\lambda))-\ln 2]}\nonumber\\
&\exe&\exp\left\{n\max_{\delta(R_1)\le\delta_1\le 1-\delta(R_1)}\left[R+\lambda h(\delta_1)
+(1-\lambda)h\left(\frac{\delta-\lambda\delta_1}{1-\lambda}\right)-\ln 2\right]\right\}.
\end{eqnarray}
Consequently, the exponential growth rate of $N_2(\delta)$ is given by
$$S(\delta)=\max_{\delta(R_1)\le\delta_1\le 1-\delta(R_1)}\left[R+\lambda h(\delta_1)
+(1-\lambda)h\left(\frac{\delta-\lambda\delta_1}{1-\lambda}\right)-\ln 2\right].$$
For large $\delta$, the constraint $\delta(R_1)\le\delta_1\le 1-\delta(R_1)$
is inactive and the achiever of $S(\delta)$ is $\delta_1=\delta$, and then
$$S(\delta)=
R+\lambda h(\delta)+(1-\lambda)h(\delta)-\ln 2=R+h(\delta)-\ln 2.$$
If we now gradually reduce $\delta$, the behavior depends on whether 
we first encounter the value $\delta=\delta(R_1)$,
below which $\delta_1=\delta$ no longer satisfies the constraint, or the 
the value $\delta=\delta(R)$, below which 
$S(\delta)=R+h(\delta)-\ln 2$
vanishes. This in turn depends on whether $\delta(R_1)$ is larger or smaller than $\delta(R)$,
or equivalently, if $R_1< R< R_2$ or $R_1\ge R\ge R_2$.

Consider the case $R_1\ge R\ge R_2$ first. In this case, $\delta(R_1)\le\delta(R)\le \delta(R_2)$, and we have:
\begin{equation}
S(\delta)=\left\{\begin{array}{ll}
R+h(\delta)-\ln 2 & \delta(R)<\delta < 1-\delta(R)\\
0 & \delta=\delta(R)~~\mbox{or}~~\delta = 1-\delta(R)\\
-\infty & \delta<\delta(R)~~\mbox{or}~~\delta > 1-\delta(R)
\end{array}\right.
\end{equation}
exactly like in the ordinary, non--hierarchical ensemble (cf.\ eq.\ (\ref{srem})), and
then the corresponding exponential rate of the partition function is as in
Subsection \ref{remc}, except that here $\beta$ is not scaled by $B$, i.e.,
$\phi(\beta)=-u(\beta,R)$.

The other case is $R_1< R< R_2$, which is equivalent to $\delta(R_1)>\delta(R)> \delta(R_2)$.
Here, in analogy to the GREM with two phase transitions, we have:
$$\phi(\beta)=\left\{\begin{array}{ll}
-v(\beta,R) & \beta < \beta(R_1)\\
-\lambda\beta\delta(R_1)-(1-\lambda)v(\beta,R_2)
& \beta(R_1)\le \beta<\beta(R_2)\\
-\beta[\lambda\delta(R_1)+(1-\lambda)\delta(R_2)] & \beta > \beta(R_2)
\end{array}\right.$$
We now identify the first line as the purely paramagnetic phase,
the second line -- as the ``semi--glassy'' phase (where $\{\hbx\}$ are glassy but $\{\tbx\}$
are paramagnetic), and the third line -- as the purely glassy phase.
Note that the glassy phase here behaves as if the two parts of the code, 
at rates $R_1$ and $R_2$, were operating independently, namely, 
as if $\{\calC_2(i)\}_{i=1}^{M_1}$ were all identical,
in which case, the distortion would have been minimized separately over the
two segments. We will get back to this point in the sequel.

We have seen then that the ensemble behaves substantially differently
depending on whether $R_1\ge R_2$ or $R_1 < R_2$. In the former case,
the above calculation may indicate that the ensemble performance is similar
to that of an ordinary block code of length $n$ without any structure.
We next carry out a detailed analysis of the characteristic function
and its exponential rate, which we shall denote by $\psi(s,R_1,R_2)$.

Similarly as before, we first compute $\bE\{Z^{1/\theta}(s\theta|\bX)\}$:
\begin{eqnarray}
\bE\{Z^{1/\theta}(s\theta|\bX)\}&=&\bE\left\{\left[\sum_{d_1=0}^{n_1}\sum_{d_2=0}^{n_2}
N(d_1,d_2)\cdot e^{-s\theta(d_1+d_2)}\right]^{1/\theta}\right\}\nonumber\\
&\exe&\sum_{d_1=0}^{n_1}\sum_{d_2=0}^{n_2}\bE\{N^{1/\theta}(d_1,d_2)\}\cdot e^{-s(d_1+d_2)},
\end{eqnarray}
where $N(d_1,d_2)$ is the number concatenated codewords $\{(\hbx,\tbx)\}$ for which the
first stage contributes distance $d_1$ and the second
stage contributes distance $d_2$. For the moments $\bE\{N^{1/\theta}(d_1,d_2)\}$, 
or equivalently, $\bE\{N^{1/\theta}(n_1\delta_1,n_2\delta_2)\}$, the following is
proven in Section A.2 of the Appendix:
\begin{equation}
\label{nd1d2}
\bE\{N^{1/\theta}(n_1\delta_1,n_2\delta_2)\}\exe\left\{\begin{array}{ll}
\exp\{n[\lambda W_1+(1-\lambda)W_2]\} & 
\delta_1 \in\calI^c(R_1),~\delta_2\in\calI^c(R_2)\\
\exp\{n[\lambda W_1+
(1-\lambda)W_2/\theta]\} & \delta_1\in\calI^c(R_1),~\delta_2\in\calI(R_2)\\
\exp\{n[\lambda W_1+(1-\lambda)W_2]/\theta\} &
\delta_1 \in\calI(R_1),~\delta_2\in\calI(R_2)\\
\exp\{n\eta[\lambda W_1+(1-\lambda)W_2]\} & 
\delta_1 \in\calI(R_1),~\delta_2\in\calI^c(R_2)
\end{array}\right.
\end{equation}
where $\calI(R)\dfn(\delta(R),1-\delta(R))$, $\calI^c(R)=[0,1]\setminus\calI(R)$,
$W_i=W(\delta_i,R_i)$, $i=1,2$, with $W(\delta,R)$ being defined as
$$W(\delta,R)\dfn R+h(\delta)-\ln 2$$ 
and
$$\eta=\eta(\theta,\delta_1,\delta_2,\lambda,R)=\left\{\begin{array}{ll}
1 & \lambda W_1+(1-\lambda)W_2< 0\\
\frac{1}{\theta} & \lambda W_1+(1-\lambda)W_2\ge 0
\end{array}\right.$$

Therefore,
\begin{eqnarray}
\bE\{Z^{1/\theta}(s\theta|\bX)\}&\exe&
\sum_{\delta_1\in\calI^c(R_1)}\sum_{\delta_2\in\calI^c(R_2)}
e^{n[R+\lambda h(\delta_1)+(1-\lambda)h(\delta_2)-\ln 2]}\times\nonumber\\
& &e^{-sn[\lambda\delta_1+(1-\lambda)\delta_2]}+\nonumber\\
& &\sum_{\delta_1\in\calI^c(R_1)}\sum_{\delta_2\in\calI(R_2)}
e^{n[\lambda(R_1+h(\delta_1)-\ln 2)+(1-\lambda)(R_2+h(\delta_2)-\ln 2)/\theta]}\times\nonumber\\
& &e^{-sn[\lambda\delta_1+(1-\lambda)\delta_2]}+\nonumber\\
& &\sum_{\delta_1\in\calI(R_1)}\sum_{\delta_2\in\calI(R_2)}
e^{n[\lambda(R_1+h(\delta_1)-\ln 2)+(1-\lambda)(R_2+h(\delta_2)-\ln 2)]/\theta}\times\nonumber\\
& &e^{-sn[\lambda\delta_1+(1-\lambda)\delta_2]}+\nonumber\\
& &\sum_{\delta_1\in\calI(R_1)}\sum_{\delta_2\in\calI^c(R_2)}
e^{n\eta[\lambda(R_1+h(\delta_1)-\ln 2)+(1-\lambda)(R_2+h(\delta_2)-\ln 2)]}\times\nonumber\\
& &e^{-sn[\lambda\delta_1+(1-\lambda)\delta_2]}\nonumber\\
&\dfn& A+B+C+D
\end{eqnarray}
Let us now handle each one of these four terms and take the limit $\theta\to\infty$.
This results in:
\begin{eqnarray}
A&\exe&\left[\sum_{\delta_1\in\calI^c(R_1)}e^{n_1[R_1+h(\delta_1)-\ln 2-s\delta_1]}\right]\cdot
\left[\sum_{\delta_2\in\calI^c(R_2)}e^{n_2[R_2+h(\delta_2)-\ln 2-s\delta_2]}\right]\nonumber\\
&\exe&e^{-n_1u(s,R_1)}\cdot e^{-n_2u(s,R_2)}\nonumber\\
&=& e^{-n[\lambda u(s,R_1)+(1-\lambda)u(s,R_2)]},
\end{eqnarray}
\begin{eqnarray}
B&\exe&\left[\sum_{\delta_1\in\calI^c(R_1)}e^{n_1[R_1+h(\delta_1)-\ln 2-s\delta_1]}\right]\cdot
\left[\sum_{\delta_2\in\calI(R_2)}e^{-n_2s\delta_2}\right]\nonumber\\
&\exe&e^{-n_1u(s,R_1)}\cdot e^{-n_2\delta(R_2)}\nonumber\\
&=& e^{-n[\lambda u(s,R_1)+(1-\lambda)\delta(R_2)]},
\end{eqnarray}
\begin{equation}
C\exe e^{-n[\lambda\delta(R_1)+(1-\lambda)\delta(R_2)]},
\end{equation}
and
\begin{equation}
D\exe e^{-nf(s,R_1,R_2)}
\end{equation}
where
$$f(s,R_1,R_2)=\min_{\delta_1\in\calI(R_1),\delta_2\in\calI^c(R_2)}
\{s[\lambda\delta_1+(1-\lambda)\delta_2]-
\mu(\delta_1,\delta_2)[R+\lambda h(\delta_1)+(1-\lambda)h(\delta_2)-\ln 2]\}$$
and where
$$\mu(\delta_1,\delta_2)=\left\{\begin{array}{ll}
1 & R+\lambda h(\delta_1)+(1-\lambda)h(\delta_2)<\ln 2\\
0 & R+\lambda h(\delta_1)+(1-\lambda)h(\delta_2)\ge\ln 2 \end{array}\right.$$
Among the terms $A$, $B$, and $C$, the term $A$ is exponentially the dominant one.
To check whether or not $A$ dominates also $D$, we will have to investigate the function
$f(s,R_1,R_2)$. This is done in Subsection A.3 of the Appendix, where it is
shown that
this function is as follows:
For $R_1 > R_2$:
\begin{equation}
\label{fr1gr2}
f(s,R_1,R_2)=\left\{\begin{array}{ll}
u(s,R) & 0\le s\le s_{R_1}\\
\lambda s\delta(R_1)+(1-\lambda)v(s,R_2) & s > s_{R_1}
\end{array}\right.
\end{equation}
and for $R_1 < R_2$:
\begin{equation}
\label{fr1lr2}
f(s,R_1,R_2)=\left\{\begin{array}{ll}
s[\lambda\delta(R_1)+(1-\lambda)\delta(R_2)] & 0\le s\le s_{R_2}\\
\lambda s\delta(R_1)+(1-\lambda)v(s,R_2) & s > s_{R_2}\end{array}\right.
\end{equation}
Finally, the overall exponential rate of the characteristic function, $\psi(s,R_1,R_2))$, we have to take
into account the contribution of $A$, as mentioned above. This gives:
$$\psi(s,R_1,R_2))=\min\{f(s,R_1,R_2),a(s,R_1,R_2)\}$$
where $a(s,R_1,R_2)\dfn \lambda u(s,R_1)+(1-\lambda)u(s,R_2)$.
Now, in the case $R_1> R_2$, for small $s$, the function $f$ is linear
with slope $\delta(R)$, whereas the function $a$
is linear with a slope of $\lambda\delta(R_1)+(1-\lambda)\delta(R_2)$ which is
larger. Thus, $f$ is smaller in some interval of small $s$. However, for
larger $s$, $f$ continues to have a linear term with slope $\lambda \delta(R_1)$ whereas
$a$ never exceeds the level of $\ln 2-R$.
Thus, there must be a (unique) point of intersection $s^*$. Consequently,
for $R_1> R_2$, we have
$$\psi(s,R_1,R_2)=\left\{\begin{array}{ll}
f(s,R_1,R_2) & s\le s^*\\
a(s,R_1,R_2) & s\ge s^*
\end{array}\right.$$
where $f(s,R_1,R_2)$ is as in (\ref{fr1gr2}).
Concerning the case $R_1 < R_2$, both $f$ (of eq.\ (\ref{fr1lr2})) and $a$
start as linear functions of the 
same slope of $\lambda\delta(R_1)+(1-\lambda)\delta(R_2)$.
However, while the latter begins its 
curvy part at $s=s_{R_1}$, the former continues
to be linear until the point $s=s_{R_2}> s_{R_1}$. In this case, 
then it is easy to see that $\psi(s,R_1,R_2)$ is dominated by $a$ across
the entire range $s\ge 0$, i.e., 
$$\psi(s,R_1,R_2)=\lambda u(s,R_1)+(1-\lambda)u(s,R_2).$$

We see then that the ensemble performance
is substantially different in the two cases: For $R_1 < R_2$,
$\psi(s,R_1,R_2)$ is exactly the same as if we used two {\it independent}
block codes of lengths $n_1$ and $n_2$
at rates $R_1$ and $R_2$, respectively. 
In particular, the corresponding average 
distortion is $\lambda\delta(R_1)+(1-\lambda)\delta(R_2)$ which
is, of course, larger than $\delta(R)$.
In other words, we are 
gaining nothing from the tree structure and the dependence between the two parts of the code. 
For $R_1 > R_2$,
on the other hand, there is at least a considerable range of small $s$ for
which $\psi(s,R_1,R_2)=u(s,R)$, namely,
the ensemble performance is exactly like that of the
ordinary ensemble of full block code 
of length $n$ and rate $R$, without any structure (which is also the best achievable exponential rate).
However, beyond a certain value of $s$, there is some loss
in comparison to the ordinary ensemble. 
The case $R_1=R_2=R$ can be obtained as the limiting behavior of both
$R_1<R_2$ and $R_1> R_2$, by taking both rates to be arbitrarily close
to each other. In this case, we obtain $\psi(s,R_1,R_2)=u(s,R)$ throughout the
{\it entire} range $s\ge 0$ (cf.\ the discussion on this in the Introduction).
The conclusion then is that
if we use an hierarchical structure of the kind we consider in this paper,
it is best to assign equal rates at the two stages, but then we might as
well abandon the tree structure of the code altogether, and just encode the two parts
independently, both at rate $R$ (this will moreover save complexity at the encoder).
If, however, certain considerations
dictate different rates at different segments, then it is better to 
encode at a larger rate in the first segment and at a smaller rate in the
second. 

This derivation can be extended, in principle, to any finite
number $k$ of stages. The analysis is, of course, more complicated but
conceptually, the ideas are the same. We will not carry out this extension
in this paper.

\subsection{Channel Coding}
\label{channelcoding}

In complete duality to the source coding problem, one may consider
a channel code (for the BSC)
with a similar hierarchical structure: Given a binary
information vector of length $nR=n_1R_1+n_2R_2$ nats, we encode it in two
parts: The first segment, of length $n_1R_1$ nats, is encoded
to a binary channel input vector of length $n_1$, independently of the forthcoming
$n_2R_2$ nats (thus, the channel encoder is of reduced delay). Then, the
remaining $n_2R_2$ nats are mapped to another binary channel input vector of length
$n_2$ and it depends on the entire information vector of length $nR$.

The ensemble of codebooks is drawn similarly as before: first, a
randomly drawn first--stage codebook of size $e^{n_1R_1}$, and then, for each one of its
codewords, another codebook of size $e^{n_2R_2}$ 
is drawn independently. Once again, each bit of
each codeword is drawn by independent fair coin tossing.

The decoder applies maximum likelihood (ML) decoding based on the
entire channel output vector $\by$ of length $n=n_1+n_2$, pertaining to 
the input $\bx$ of length $n$. The analogy with the GREM is that here,
the energy function is the log--likelihood, which is additive over
the two stages by the memorylessness of the channel.

In full analogy to the GREM and the 
source coding problem of Subsection \ref{lossysourcecoding}, and as an extension
to the derivation in Subsection \ref{remc}, here too, the partition
function $Z_e(\beta|\by)$ has exactly the same two different
types of behavior, depending on whether $R_1\ge R_2$
or $R_1< R_2$. Therefore, we will not repeat this here.

Concerning the aspect of performance evaluation of this ensemble of codes,
and a comparison to the ordinary ensemble, here 
the natural figure of merit is Gallager's random coding error exponent,
which can be analyzed using methods similar to those that we used
in Subsection \ref{lossysourcecoding}. We will not carry out a very refined analysis
as we did before, but we will make a few observations in this context,
although not quite directly related to the GREM.

Referring to the notation of Subsection \ref{remc}, let $\calC=
\{\bx_1,\ldots,\bx_M\}$ be a given channel code of
size $M=e^{nR}$ and block length $n$, and let $\by$ designate
the output vector of the BSC, of length $n$.
Gallager's classical upper bound \cite[p.\ 65, eq.\ (2.4.8)]{VO79} on the probability
of error is well known to be given by
$$P_e \le \frac{1}{M}\sum_{m=1}^M\sum_{\by} P(\by|\bx_m)^{1/(1+\rho)}\cdot
\left[\sum_{m'\ne m} P(\by|\bx_{m'})^{1/(1+\rho)}\right]^\rho 
~~~0\le\rho\le 1.$$
Consider first the ordinary ensemble, where all $M$ codewords are
chosen independently at random. In this case, taking the expectation
of both sides, the average
error probability is upper bounded by
$$\bar{P}_e \le \frac{1}{M}\sum_{m=1}^M\sum_{\by} 
\bE\{P(\by|\bX_m)^{1/(1+\rho)}\}\cdot
\bE\left\{\left[\sum_{m'\ne m} 
P(\by|\bX_{m'})^{1/(1+\rho)}\right]^\rho\right\}.$$
As is shown in \cite{Merhav07}, the second factor of the summand
is actually the expectation of the $\rho$--th moment of the
partition function $Z_e(\beta|\by)$ computed at the inverse temperature
$\beta=1/(1+\rho)$. Now, at least for the ordinary ensemble,
the traditional derivation, which is based on applying Jensen's inequality,
is good enough to yield an exponentially tight bound \cite{Gallager73} on the ensemble performance.
This amounts to inserting the expectation into the square brackets, i.e.,
$$\bar{P}_e \le \frac{1}{M}\sum_{m=1}^M\sum_{\by} 
\bE\{P(\by|\bX_m)^{1/(1+\rho)}\}\cdot
\left[\sum_{m'\ne m}\bE\{
P(\by|\bX_{m'})^{1/(1+\rho)}\}\right]^\rho.$$
We shall not continue any further with the analysis of this expression.
Instead, we shall compare it as is, with a corresponding upper bound
for the hierarchical ensemble defined above. 

In the hierarchical case with $k=2$ stages, the probability of error consists
of two contributions. The first pertains to all incorrect
codewords $\bx=(\bx',\bx'')$ whose first segment $\bx'$ agrees
with that of the correct codeword, and the second one is associated
with all other incorrect codewords. As for the former type of codewords,
the ML decoder actually compares the likelihood scores of the second segment
only (as those of the first segment are the same and hence cancel out), 
and so, these incorrect codewords contribute a term of the order
of $e^{-n_2E_r(R_2)}$ to the average error probability, where $E_r(R)$
is the Gallager's random coding error exponent 
function \cite[p.\ 139, eq.\ (5.6.16)]{Gallager68}. Concerning the second
set of incorrect codewords, we can apply an upper bound as above, except
that the expectations have to be taken w.r.t.\ the hierarchical ensemble.
However, it is easy to see that the expectation of 
$\bE\{P(\by|\bX)^{1/(1+\rho)}\}$ is exactly the same as in the ordinary
ensemble, and thus, so is the upper bound 
for this set of codewords, which is then
$e^{-nE_r(R)}$. The total average error probability is then
upper bounded by
$$\bar{P}_e\le e^{-nE_r(R)}+e^{-n_2E_r(R_2)}
= e^{-nE_r(R)}+e^{-n(1-\lambda)E_r(R_2)}.$$
This gives further motivation why $R_2$ should be chosen smaller than $R_1$:
If $R_2> R_1$, the second term definitely dominates the exponent, because
both $n_2 < n$ and $R_2> R$ and so $E_r(R_2) < E_r(R)$. For a given $R$ and $\lambda$, can we, and if so
how, assign the segmental rates $R_1$ and $R_2$ such that the second term
would not be dominant, i.e.,
$(1-\lambda)E_r(R_2)\ge E_r(R)$? If $R$ is large enough this is
possible. For example, one way to do this is to select $R_1=C$, where $C$ is the channel
capacity. In this case, we have, by the convexity of $E_r(\cdot)$:
$$E_r(R)=E_r(\lambda C+(1-\lambda)R_2)\le
\lambda E_r(C)+(1-\lambda)E_r(R_2)=(1-\lambda)E_r(R_2).$$
For this strategy to be applicable, $R$ must be at least as large as $\lambda C$.

How does this discussion extend to a general number of stages $k$
and is there a more systematic approach to allocate the segmental rates $R_1,\ldots,R_k$
for a given overall rate $R$?
For simplicity, let us suppose that the segment lengths
are all the same, i.e., $n_1=n_2=\ldots=n_k=n/k$.
The extension turns out to be quite straightforward: In the case of $k$ stages there
are $k$ types of incorrect codewords: Those that agree with the correct
codeword in all stages except the last stage, those that agree in all stages
except the last two stages, etc. Accordingly, using the same
considerations as above, it is easy to see then that the 
upper bound on the average error
probability consists of $k$ contributions whose exponents are
$$\frac{k-i}{k}E_r\left(\frac{1}{k-i}\sum_{j=i+1}^k R_j\right), ~~i=0,1,\ldots,k-1.$$
For convenience, let us denote 
$$\bar{R}_i=\frac{1}{k-i}\sum_{j=i+1}^k R_j.$$
Under what conditions and how can we assign the segmental rates such that
$$\frac{k-i}{k}E_r(\bar{R}_k)\ge E_r(R)$$
for all $i=1,2,\ldots,k-1$?
First, we must select $\bar{R}_1$ sufficiently small such that
$E_r(\bar{R}_1)\ge \frac{k}{k-1}E_r(R)$. 
As $R$ is given, this will dictate the choice of 
$R_1$ according to the identity
$$R=\bar{R}_0=\frac{1}{k}R_1+\frac{k-1}{k}\bar{R}_1.$$
Next, we choose $\bar{R}_2$ small enough such that 
$$E_r(\bar{R}_2)\ge \frac{k}{k-2}E_r(R).$$ 
As $\bar{R}_1$ has already been
chosen, this will dictate the choice of $R_2$ according to the identity
$$\bar{R}_1=\frac{1}{k-1}R_2+\frac{k-2}{k-1}\bar{R}_2,$$
and so on. This procedure continues until in the last step
we choose 
$R_k=\bar{R}_{k-1}$ such that $E_r(R_k)\ge kE_r(R)$, which
dictates the choice of $R_{k-1}$ via $\bar{R}_{k-2}=(R_k+R_{k-1})/2$,
where $\bar{R}_{k-2}$ was selected in preceeding step.
An obvious condition for this procedure to be applicable is that
$R$ would be 
large enough such that $E_r(R)\le E_r(0)/k$. Note that if some of the
segmental rates exceed capacity (or even the log alphabet size), this
is not a problem, as long as the averages $\bar{R}_i$ are all small enough.

\section*{Appendix}
\renewcommand{\theequation}{A.\arabic{equation}}
    \setcounter{equation}{0}

\subsection*{A.1 Proof of Eq.\ (\ref{moments})}
\label{moments1level}

We begin with a simple large deviations bound regarding
the distance enumerator, which appears also in \cite{Merhav07b},
but we present here too for the sake of
completeness.
For $a,b\in[0,1]$, consider the binary divergence
\begin{eqnarray}
D(a\|b)&\dfn&a\ln \frac{a}{b}+(1-a)\ln\frac{1-a}{1-b}\nonumber\\
&=&a\ln \frac{a}{b}+(1-a)\ln\left[1+\frac{b-a}{1-b}\right].
\end{eqnarray}
To derive a lower bound to $D(a\|b)$, let us use the inequality
\begin{equation}
\label{lnineq}
\ln(1+x)=-\ln\frac{1}{1+x}=-\ln\left(1-\frac{x}{1+x}\right)\ge \frac{x}{1+x},
\end{equation}
and then
\begin{eqnarray}
D(a\|b)&\ge&a\ln \frac{a}{b}+(1-a)\cdot\frac{(b-a)/(1-b)}
{1+(b-a)/(1-b)}\nonumber\\
&=&a\ln \frac{a}{b}+b-a\nonumber\\
&>&a\left(\ln\frac{a}{b}-1\right).
\end{eqnarray}
For every given $\by$, $N(d)$ is the sum of the $e^{nR}-1$
independent binary random variables, $\{1\{d(\bX_{m'},\by)=d\}\}_{m'\ne m}$,
where the probability that $d(\bX_{m'},\by)=n\delta$ is
exponentially $b\exe e^{-n[\ln 2- h(\delta)]}$.
The event $N(n\delta)\ge e^{nA}$, for $A\in[0,R)$,
means that the relative frequency of the event $1\{d(\bX_{m'},\by)=n\delta\}$
is at least $a=e^{-n(R-A)}$. Thus, by the Chernoff bound:
\begin{eqnarray}
\mbox{Pr}\{N(n\delta)\ge e^{nA}\}&\lexe&
\exp\left\{-(e^{nR}-1)D(e^{-n(R-A)}\|e^{-n[\ln 2-
h(\delta)]})\right\}\nonumber\\
&\lexe&
\exp\left\{-e^{nR}\cdot e^{-n(R-A)}(n[(\ln 2-R-
h(\delta)+A]-1)\right\}\nonumber\\
&\le&
\exp\left\{-e^{nA}(n[\ln 2-R-h(\delta)+A]-1)\right\}.
\end{eqnarray}
Denoting by $\calI(R)$ the interval $(\delta(R),1-\delta(R))$ and by $\calI^c(R)$,
the complementary range $[0,1]\setminus\calI(R)$,
we have, for $\delta\in\calI^c(R)$:
\begin{eqnarray}
\bE\{N^s(n\delta)\}&\le&e^{n\epsilon s}\cdot\mbox{Pr}\{1\le N(n\delta)\le e^{n\epsilon}\}+
e^{nRs}\cdot \mbox{Pr}\{ N(n\delta)\ge e^{n\epsilon}\}\nonumber\\
&\le&e^{n\epsilon s}\cdot\mbox{Pr}\{N(n\delta)\ge 1\}+
e^{nRs}\cdot \mbox{Pr}\{ N(n\delta)\ge e^{n\epsilon}\}\nonumber\\
&\le&e^{n\epsilon s}\cdot\bE\{N(n\delta)\}+
e^{nRs}\cdot e^{-(n\epsilon-1)e^{n\epsilon}}\nonumber\\
&\le&e^{n\epsilon s}\cdot e^{n[R+h(\delta)-\ln 2]}+
e^{nRs}\cdot e^{-(n\epsilon-1)e^{n\epsilon}}.
\end{eqnarray}
One can let $\epsilon$ vanish with $n$ sufficiently slowly that
the second term is still superexponentially small, e.g., $\epsilon=1/\sqrt{n}$.
Thus, for $\delta\in\calI^c(R)$, $\bE\{N^s(n\delta)\}$ is exponentially bounded
by $e^{n[R+h(\delta)-\ln 2]}$ independently of $s$. For $\delta\in\calI(R)$, we have:
\begin{eqnarray}
\bE\{N^s(n\delta)\}&\le&e^{ns[R+h(\delta)-\ln 2+\epsilon]}\cdot
\mbox{Pr}\{N(n\delta)\le e^{n[R+h(\delta)-\ln 2+\epsilon]}\}+\nonumber\\
& &e^{nRs}\cdot
\mbox{Pr}\{N(n\delta)\ge e^{n[R+h(\delta)-\ln 2+\epsilon]}\}\nonumber\\
&\le&e^{ns[R+h(\delta)-\ln 2+\epsilon]}
+e^{nRs}\cdot e^{-(n\epsilon-1)e^{n\epsilon}}
\end{eqnarray}
where again, the second term is exponentially negligible.

To see that both bounds are exponentially tight, consider
the following lower bounds. For $\delta\in\calI^c(R)$,
\begin{eqnarray}
\bE\{N^s(n\delta)\}&\ge&1^s\cdot\mbox{Pr}\{N(n\delta)=1\}\nonumber\\
&=&e^{nR}\cdot\mbox{Pr}\{d_H(\bX,\by)=n\delta\}\cdot
\left[1-\mbox{Pr}\{d_H(\bX,\by)=n\delta\}\right]^{e^{nR}-1}\nonumber\\
&\exe&e^{nR}e^{-n[\ln 2-h(\delta)]}\cdot
\left[1-e^{-n[\ln 2-h(\delta)]}\right]^{e^{nR}}\nonumber\\
&=&e^{n[R+h(\delta)-\ln 2]}\cdot\exp\{e^{nR}\ln[1-e^{-n[\ln 2-h(\delta)]}]\}.
\end{eqnarray}
Using again the inequality in (\ref{lnineq}),
the second factor is lower bounded by
$$\exp\{-e^{nR}e^{-n[\ln 2-h(\delta)]}/(1-e^{-n[\ln 2-h(\delta)]})\}
=\exp\{-e^{-n[\ln 2-R-h(\delta)]}/(1-e^{-n[\ln 2-h(\delta)]})\}$$
which clearly tends to unity as $\ln 2-R-h(\delta) > 0$ for $\delta\in\calI^c(R)$.
Thus, $\bE\{N^s(n\delta)\}$ is exponentially lower bounded by $e^{n[R+h(\delta)-\ln 2]}$.
For $\delta\in\calI(R)$, and an arbitrarily small $\epsilon > 0$, we have:
\begin{eqnarray}
\bE\{N^s(n\delta)\}&\ge& e^{ns[R+h(\delta)-\ln 2-\epsilon]}\cdot
\mbox{Pr}\{N(n\delta)\ge e^{n[R+h(\delta)-\ln 2-\epsilon]}\}\nonumber\\
&=& e^{ns[R+h(\delta)-\ln 2-\epsilon]}\cdot\left(1-
\mbox{Pr}\{N(n\delta)< e^{n[R+h(\delta)-\ln 2-\epsilon]}\}\right)
\end{eqnarray}
where $\mbox{Pr}\{N(n\delta)< e^{n[R+h(\delta)-\ln 2-\epsilon]}\}$ is again
upper bounded, for an internal point in $\calI(R)$,
by a double exponentially small quantity as above.
For $\delta$ near the boundary of $\calI(R)$, namely, when $R+h(\delta)-\ln 2\approx 0$,
we can lower bound $\bE\{N^s(n\delta)\}$ by slightly reducing $R$ to $R'=R-\epsilon$
(where $\epsilon > 0$ is very small). This will make
$\delta$ an internal point of $\calI^c(R')$ for which the previous bound applies,
and this bound
is of the exponential order of $e^{n[R'+h(\delta)-\ln 2]}$.
Since $R'+h(\delta)-\ln 2$ is still very close to zero, then $e^{n[R'+h(\delta)-\ln 2]}$
is of the same exponential order as $e^{ns[R+h(\delta)-\ln 2]}$
since both are about $e^{0\cdot n}$.

It should be noted that a similar double--exponential bound can be obtained for
the probability of the event $\{N(n\delta) \le e^{nA}\}$, where $A < R+h(\delta)-\ln 2$
and $R+h(\delta)-\ln 2> 0$.
Here we can proceed as above except that the in the lower bound on divergence $D(a\|b)$
we should take the second line of (A.3) (rather than the third), 
which is of the exponential order of $b\exe e^{-n[\ln 2-h(\delta)]}$ (observe that
here $b$ is exponentially larger than $a$, as opposed to the earlier case). Thus,
we obtain $R+h(\delta)-\ln 2>0 $ at the second level exponent, and so the decay is double
exponential as before.

\subsection*{A.2. Proof of Eq.\ (\ref{nd1d2})}
\label{moments2levels}

First, let us write $N(n_1\delta_1,n_2\delta_2)$ as follows:
\begin{eqnarray}
N(n_1\delta_1,n_2\delta_2)&=&\sum_{i=1}^{M_1}1\{d_H(\bx',\hbx_i)=n_1\delta_1\}\cdot
\sum_{j=1}^{M_2}1\{d_H(\bx'',\tbx_{i,j})=n_2\delta_2\}\nonumber\\
&\dfn&\sum_{i=1}^{M_1}1\{d_H(\bx',\hbx_i)=n_1\delta_1\}\cdot N_i(n_2\delta_2)
\end{eqnarray}
where $\bx'$ and $\bx''$ designate $(x_1,\ldots,x_{n_1})$ and $(x_{n_1+1},\ldots,x_n)$,
respectively, and where $1\{\cdot\}$ denotes the indicator function of an event.
We now treat each one of the four cases pertaining to the combinations of both
$\delta_1$ and $\delta_2$ being or not being members of $\calI(R_1)$ and $\calI(R_2)$,
respectively.\\

\subsubsection*{Case 1: $\delta_1\in\calI^c(R_1)$ and $\delta_2\in\calI^c(R_2)$}
\label{case1}

For a given, arbitrarily small
$\epsilon > 0$, consider the event $\calE=\{N(n_1\delta_1,n_2\delta_2)\ge e^{n\epsilon}\}$.
If both the
number of indices $i$ for which $d_H(\bx',\hbx_i)=n_1\delta_1$ is less than $e^{n_1\epsilon}$
and for each $i$,
$N_i(n_2\delta_2)\le e^{n_2\epsilon}$, then clearly, the event $\calE$ does not occur.
Thus, for $\calE$ to occur, at least one of these events must occur. In other words,
either the number of indices $i$ for which $d_H(\bx',\hbx_i)=n_1\delta_1$ is
larger than $e^{n_1\epsilon}$ or there exist $i$ for which
$N_i(n_2\delta_2)> e^{n_2\epsilon}$. The probability of the former
event is upper bounded by
$e^{-e^{n_1\epsilon}(n_1\epsilon-1)}$ (cf.\ Subsection A.1). Similarly,
the probability of the latter, for a given $i$, is bounded by $e^{-e^{n_2\epsilon}(n_2\epsilon-1)}$. Thus,
the probability of the union of events $\bigcup_i\{N_i(n_2\delta_2)>
e^{n_2\epsilon}\}$ is upper bounded by
$M_1e^{-e^{n_2\epsilon}(n_2\epsilon-1)}=e^{n_1R_1}\cdot e^{-e^{n_2\epsilon}(n_2\epsilon-1)}$, which
is still double exponential in $n$. Thus,
$$\mbox{Pr}\{\calE\}\le e^{-e^{n_1\epsilon}(n_1\epsilon-1)}+e^{n_1R_1}\cdot
e^{-e^{n_2\epsilon}(n_2\epsilon-1)}.$$
Therefore,
\begin{eqnarray}
\bE\{N^{1/\theta}(n_1\delta_1,n_2\delta_2)\}&\le& 
0^{1/\theta}\cdot\mbox{Pr}\{N(n_1\delta_1,n_2\delta_2)=0\}+
e^{n\epsilon/\theta}\cdot\mbox{Pr}\{1\le N(n_1\delta_1,n_2\delta_2)\le e^{n\epsilon}\}\nonumber\\
& &+e^{nR/\theta}\cdot\mbox{Pr}\{\calE\}\nonumber\\
&\le&e^{n\epsilon/\theta}\cdot\mbox{Pr}\{N(n_1\delta_1,n_2\delta_2)\ge 1\}+
e^{nR/\theta}\cdot\mbox{Pr}\{\calE\}\nonumber\\
&\le&e^{n\epsilon/\theta}\cdot\bE\{N(n_1\delta_1,n_2\delta_2)\}+
e^{nR/\theta}\cdot\mbox{Pr}\{\calE\},
\end{eqnarray}
which is exponentially upper bounded by $e^{n[R+\lambda h(\delta_1)+(1-\lambda)h(\delta_2)-\ln 2]}$
since $\epsilon$ is arbitrarily small, $\bE\{N((n_1\delta_1,n_2\delta_2)\}\exe
e^{n[R+\lambda h(\delta_1)+(1-\lambda)h(\delta_2)-\ln 2]}$, and the last term is
double--exponential. To obtain the compatible lower bound, we use 
\begin{eqnarray}
\bE\{N^{1/\theta}(n_1\delta_1,n_2\delta_2)\}&\ge&
1^{1/\theta}\cdot\mbox{Pr}\{N(n_1\delta_1,n_2\delta_2)=1\}\nonumber\\
&=&\mbox{Pr}\{N(n_1\delta_1,n_2\delta_2)=1\}.
\end{eqnarray}
Now, the event $\{N(n_1\delta_1,n_2\delta_2)=1\}$ is the event that there is exactly
one value of $i$ such that $d_H(\bx',\hbx)=n_1\delta_1$, and that for this $i$, there is
exactly one $j$ such that $d_H(\bx'',\tbx)=n_2\delta_2$. As shown in Subsection A.1,
the probability of the former is exponentially $e^{n_1[R_1+h(\delta_1)-\ln 2]}$ and
the probability of the latter is exponentially $e^{n_2[R_2+h(\delta_2)-\ln 2]}$. Thus,
by independence, $\mbox{Pr}\{N(n_1\delta_1,n_2\delta_2)=1\}$ is the product, which is
exponentially $e^{n[R+\lambda h(\delta_1)+(1-\lambda)h(\delta_2)-\ln 2]}$.

\subsubsection*{Cases 2 and 3: $\delta_2\in\calI(R_2)$}
\label{cases23}

Define now the event $\calA$ as 
$$\calA=\bigcap_{i=1}^{M_1}\left\{N_i(n_2\delta_2)\le 
\exp\{n_2[R_2+h(\delta_2)-\ln 2+\epsilon]\}\right\}.$$
As we have argued before, the probability of $\calA$ is doubly exponentially
close to unity (since the probability of $\calA^c$ is upper bounded by the sum of exponentially
many doubly-exponentially small probabilities). Now, clearly, if $\calA$ occurs,
$$N(n_1\delta_1,n_2\delta_2)\le \exp\{n_2[R_2+h(\delta_2)-\ln 2+\epsilon]\}\cdot\sum_{i=1}^{M_1}
1\{d_H(\bx',\hbx_i)=n_1\delta_1\}.$$
Thus,
\begin{eqnarray}
\bE\{N^{1/\theta}(n_1\delta_1,n_2\delta_2)\}&\le&\mbox{Pr}\{\calA\}\cdot\bE\left\{\left[
\exp\{n_2[R_2+h(\delta_2)-\ln 2+\epsilon]\}\times\right.\right.\nonumber\\
& &\left.\left.\sum_{i=1}^{M_1}
1\{d_H(\bx',\hbx_i)=n_1\delta_1\}\right]^{1/\theta}\right\}\nonumber\\
& &+e^{nR/\theta}\cdot\mbox{Pr}\{\calA^c\},
\end{eqnarray}
where the second term is again doubly--exponentially small. As for the first term,
we bound $\mbox{Pr}\{\calA\}$ by unity and
\begin{eqnarray}
&&\bE\left\{\left[\exp\{n_2[R_2+h(\delta_2)-\ln 2+\epsilon]\}\sum_{i=1}^{M_1}
1\{d_H(\bx',\hbx_i)=n_1\delta_1\}\right]^{1/\theta}\right\}\nonumber\\
&=&\exp\{n_2[R_2+h(\delta_2)-\ln 2+\epsilon]/\theta\}\cdot\bE\left\{\left[\sum_{i=1}^{M_1}
1\{d_H(\bx',\hbx_i)=n_1\delta_1\}\right]^{1/\theta}\right\}
\end{eqnarray}
where the latter expectation (cf.\ Subsection A.1) is of the exponential order of 
$e^{n_1[R_1+h(\delta_1)-\ln 2]}$ if $\delta_1\in\calI^c(R_1)$ (Case 2) and 
$e^{n_1[R_1+h(\delta_1)-\ln 2]/\theta}$ if $\delta_1\in\calI(R_1)$ (Case 3).
Thus, in both cases, we obtain the desired exponential order as an upper bound.
For the lower bound, we argue similarly that the probability of the event
$$\calA'=\bigcap_{i=1}^{M_1}\left\{N_i(n_2\delta_2)\ge 
\exp\{n_2[R_2+h(\delta_2)-\ln 2-\epsilon]\}\right\}$$
is doubly--exponentially close to unity, and so,
$$\bE\{N^{1/\theta}(n_1\delta_1,n_2\delta_2)\}\ge\mbox{Pr}\{\calA\}\cdot\bE\left\{\left[
\exp\{n_2[R_2+h(\delta_2)-\ln 2-\epsilon]\}\sum_{i=1}^{M_1}
1\{d_H(\bx',\hbx_i)=n_1\delta_1\}\right]^{1/\theta}\right\},$$
and we again use the above result on the moments of 
$\sum_{i=1}^{M_1}1\{d_H(\bx',\hbx_i)=n_1\delta_1\}$ in both cases
of $\delta_1$.

\subsubsection*{Case 4: $\delta_1\in\calI(R_1)$ and $\delta_2\in\calI^c(R_2)$}
\label{case4}

Since $\delta_1\in\calI(R_1)$, then the event
$$\calA=\left\{e^{n_1[R_1+h(\delta_1)-\ln 2-\epsilon]}\le
\sum_{i=1}^{M_1}1\{d_H(\bx',\hbx_i)=n_1\delta_1\}
\le e^{n_1[R_1+h(\delta_1)-\ln 2+\epsilon]}\right\},$$
has a probability which is doubly--exponentially close to unity.
Thus, given that $\calA$ occurs, there are 
$$e^{n_1[R_1+h(\delta_1)-\ln 2+\epsilon]}\le L\le  e^{n_1[R_1+h(\delta_1)-\ln 2+\epsilon]}$$
indices $i_1,i_2,\ldots,i_L$ for which $d_H(\bx',\hbx_i)=n_1\delta_1$. Given $L$ and given these 
indices, $N(n_1\delta_1,n_2\delta_2)$ is the sum of 
$LM_2\exe e^{n_1[R_1+h(\delta_1)-\ln 2+]+n_2R_2}$ i.i.d.\ Bernoulli trials,
$1\{d_H(\bx'',\tbx)=n_2\delta_2\}$, whose probability of success is exponentially $q\exe e^{n_2[h(\delta_2)
-\ln 2]}$. Thus, similarly as in the derivation in Subsection A.1,
$$\bE\{N^{1/\theta}(n_1\delta_1,n_2\delta_2)|\calA\}\exe\left\{\begin{array}{ll}
LM_2q & q \gexe  LM_2\\
(LM_2q)^{1/\theta} & q \lexe  LM_2
\end{array}\right.$$
or, equivalently, in the notation of eq. (\ref{nd1d2}):
$$\bE\{N^{1/\theta}(n_1\delta_1,n_2\delta_2)|\calA\}\exe\left\{\begin{array}{ll}
\exp\{n[\lambda W_1+(1-\lambda)W_2]\} & \lambda W_1+(1-\lambda)W_2 < 0\\
\exp\{n[\lambda W_1+(1-\lambda)W_2]/\theta\} & \lambda W_1+(1-\lambda)W_2 \ge 0\end{array}\right.$$
The total expectation should, of course, account for $\calA^c$ as well,
but since the probability of this event is doubly exponentially small,
then the contribution of this term is negligible.

This completes the proof of eq.\ (\ref{nd1d2}).

\subsection*{A.3. The function $f(s,R_1,R_2)$}
\label{fsr1r2}

First, we observe that the constraints $\delta_1\in\calI(R_1)$
and $\delta_2\in\calI^c(R_2)$ can be replaced by their one-sided versions
$\delta_1\ge\delta(R_1)$ and $\delta_2\le\delta(R_2)$, respectively,
since values of $\delta_1$ and $\delta_2$ beyond $0.5$ cannot be better
than their corresponding reflections $1-\delta_1$ and $1-\delta_2$.

Next observe that $f(s,R_1,R_2)$ can be rewritten as follows:
$$f(s,R_1,R_2)=\min\{f_1(s,R_1,R_2)),f_2(s,R_1,R_2)\},$$
where
$$f_1(s,R_1,R_2)=s\min [\lambda\delta_1+(1-\lambda)\delta_2]$$
subject to the constraints $\delta_1\ge\delta(R_1)$, $\delta_2\le\delta(R_2)$,
and $R+\lambda h(\delta_1)+(1-\lambda)h(\delta_2)\ge \ln 2$, and
$$f_2(s,R_1,R_2)=\min\{\lambda[s\delta_1-R_1-h(\delta_1)+\ln 2]+
(1-\lambda)[s\delta_2-R_2-h(\delta_2)+\ln 2]\}$$
subject to the constraints $\delta_1\ge\delta(R_1)$, $\delta_2\le\delta(R_2)$,
and $R+\lambda h(\delta_1)+(1-\lambda)h(\delta_2)\le \ln 2$.
Note that the optimization problem associated with $f_1(s,R_1,R_2)$
is a convex problem, but the one pertaining to $f_2(s,R_1,R_2)$ is not,
because of its last constraint which is not convex.

At this point, we have to distinguish between two cases:
(i) $R_1>R_2$ and (ii) $R_2< R_1$ (the case $R_1=R_2$ will be
taken as a limit $R_1\to R_2$ of case (i)).

\subsubsection*{The Case $R_1> R_2$}
\label{r1gr2}

When $R_1> R_2$, we have $\delta(R_1)< \delta(R)<\delta(R_2)$. 
As for $f_1$, it is easy to see that $\delta_1=\delta_2=\delta(R)$
is a solution that satisfies the necessary and sufficient Kuhn--Tucker conditions
for optimality of a convex problem, and so, $f_1(s,R_1,R_2)=s\delta(R)$.

Consider next the function $f_2(s,R_1,R_2)$.
Let us ignore, for a moment, the non--convex
constraint $R+\lambda h(\delta_1)+(1-\lambda) h(\delta_2)\le 2$,
and refer only to the constraints $\delta_1\ge \delta(R_1)$
and $\delta_2\le\delta(R_2)$. Denote by $\tilde{f}_2(s,R_1,R_2)$ the
corresponding maximum without the non--convex constraint. 
The maximization problem associated with $\tilde{f}_2$
is now convex and it is to see that
$\delta_1^*=\max\{\delta(R_1),\nu_s\}$ and
$\delta_2^*=\min\{\delta(R_2),\nu_s\}$ satisfy 
the necessary and sufficient conditions for optimality,
where $\nu_s\dfn 1/(1+e^s)$.
This is also a solution for $f_2$ if it satisfies 
the non--convex constraint, namely, if
\begin{equation}
\label{cond}
\lambda h\left(\max\{\delta(R_1),\nu_s\}\right)+
(1-\lambda)h\left(\min\{\delta(R_2),\nu_s\}\right)+R \le \ln 2.
\end{equation}
Whether or not this condition is satisfied depends on $s$.
Since we are assuming $R_1> R_2$, we then have
$s_{R_1}> s_{R_2}$, where we remind that 
$s_R\dfn\ln\frac{1-\delta(R)}{\delta(R)}$.
Consequently, there are three different ranges of $s$:
$s > s_{R_1}$, $s_{R_2}<s\le s_{R_1}$, and $s\le s_{R_2}$.

When $s> s_{R_1}> s_{R_2}$, this is equivalent to
$\nu_s < \delta(R_1)<\delta(R_2)$ in which case the
above necessary condition (\ref{cond}) becomes
$$\lambda h(\delta(R_1))+
(1-\lambda)h(\nu_s) < \ln 2-R.$$
To check whether this condition is satisfied,
observe that $h(\delta(R_1))\equiv\ln 2-R_1$, and so this is equivalent
to the condition $h(\nu_s)<\ln 2-R_2$, which is $\nu_s<\delta(R_2)$,
in agreement with the assumption on the range of $s$. Therefore,
the above solution is acceptable for $f_2$ and by substituting it back into the
objective function, we get:
\begin{eqnarray}
f_2(s,R_1,R_2)&=&\lambda[s\delta(R_1)-R_1-h(\delta(R_1))+\ln 2]+
(1-\lambda)[s\nu_s-R_2-h(\nu_s)+\ln 2]\nonumber\\
&=&\lambda s\delta(R_1)+(1-\lambda)v(s,R_2)
\end{eqnarray}
When $s_{R_1}\ge s > s_{R_2}$, this is equivalent to
$\delta(R_1)<\nu_s<\delta(R_2)$, in which case 
the condition (\ref{cond}) becomes $h(\nu_s) < \ln 2-R$,
or equivalently, $\nu_s < \delta(R)$, which is $s> s_R$.
However, $s_R$ is between $s_{R_1}$ and $s_{R_2}$, and so,
the conclusion is that the
non--convex constraint is satisfied only in upper part of the interval $[s_{R_2},
s_{R_1}]$, i.e., $[s_R,s_{R_1}]$. In this range, $\delta_1^*=\delta_2^*=\nu_s$,
and this yields $f_2(s,R_1,R_2)=v(s,R)$. 
For $s < s_{R}$, the condition (\ref{cond}) no longer holds. 
In this case, the optimum solution should be sought on the boundary
of the non--convex constraint, namely, under the equality constraint
$R+\lambda h(\delta_1)+(1-\lambda) h(\delta_2)=\ln 2$, but this coincides
then with the solution to $f_1$ which was found on this boundary as well.
Thus, for $s\in[0,s_R]$, we have $f_2(s,R_1,R_2)=s\delta(R)$.
Summarizing our results for $f_2$ over the entire range of $s\ge 0$, we have
$$f_2(s,R_1,R_2)=\left\{\begin{array}{ll}
s\delta(R) & 0\le s\le s_R\\
v(s,R) & s_R<s\le s_{R_1}\\
\lambda s\delta(R_1)+(1-\lambda)v(s,R_2) & s > s_{R_1}
\end{array}\right.$$
or, equivalently,
$$f_2(s,R_1,R_2)=\left\{\begin{array}{ll}
u(s,R) & 0\le s\le s_{R_1}\\
\lambda s\delta(R_1)+(1-\lambda)v(s,R_2) & s > s_{R_1}
\end{array}\right.$$
Finally, $f$ should be taken as the minimum between $f_1$ and $f_2$.
Now, $f_1$ is linear and $f_2$ is concave (as it is the minimum of a linear
function in $s$), coinciding with $f_1$ along $[0,s_R]$. Thus $f_2$ cannot
exceed $f_1$ for any $s$, and so, $f=f_2$. Thus,
$$f(s,R_1,R_2)=\left\{\begin{array}{ll}
u(s,R) & 0\le s\le s_{R_1}\\
\lambda s\delta(R_1)+(1-\lambda)v(s,R_2) & s > s_{R_1}
\end{array}\right.$$

\subsubsection*{The Case $R_1< R_2$}
\label{r1lr2}

In this case, $\delta(R_1)>\delta(R_2)$.
Once again, $f_1$ is associated with a convex program
whose conditions for optimality are easily seen to be satisfied by
the solution $\delta_1=\delta(R_1)$ and $\delta_2=\delta(R_2)$. Thus,
$$f_1(s,R_1,R_2)=s[\lambda\delta(R_1)+(1-\lambda)\delta(R_2)].$$

As for $f_2$, let us examine again the various ranges of $s$, where 
this time, $s_{R_1}<s_R< s_{R_2}$. For $s > s_{R_2}$, we have $\nu_s<\delta(R_2)<\delta(R_1)$
and then the condition (\ref{cond}) is equivalent to $h(\nu_s)\le \ln 2-R_2$, which is
$\nu_s<\delta(R_2)$, in agreement with the assumption. This corresponds to $\delta_1=\delta(R_1)$
and $\delta_2=\nu_s$, which yields
$$f_2(s,R_1,R_2)=\lambda s\delta(R_1)+(1-\lambda)v(s,R_2).$$
For $s_{R_1}<s<s_{R_2}$, which means $\delta(R_2)<\nu_s<\delta(R_1)$, condition (\ref{cond})
is satisfied with equality, and the corresponding solution is $\delta_1=\delta(R_1)$
and $\delta_2=\delta(R_2)$, which yields
$$f_2(s,R_1,R_2)=s[\lambda \delta(R_1)+(1-\lambda)\delta(R_2)].$$
For $s < s_{R_1}$, eq.\ (\ref{cond}) is not satisfied, and we resort again to the
boundary solution, which, as mentioned earlier, is the same as $f_1$.
Summarizing our findings for the case $R_1 < R_2$, and applying
similar concavity considerations
as before (telling us that $f=f_2$), we have:
$$f(s,R_1,R_2)=\left\{\begin{array}{ll}
s[\lambda\delta(R_1)+(1-\lambda)\delta(R_2)] & 0\le s\le s_{R_2}\\
\lambda s\delta(R_1)+(1-\lambda)v(s,R_2) & s > s_{R_2}\end{array}\right.$$

\end{document}